\documentclass[a4paper,twocolumn,11pt,accepted=2023-01-30]{quantumarticle}
\pdfoutput=1
\usepackage[utf8]{inputenc}
\usepackage[english]{babel}
\usepackage[T1]{fontenc}
\usepackage{amsmath}
\usepackage{hyperref}
\usepackage{tikz}
\usepackage{lipsum}
\usepackage{epsfig}  
\usepackage{amsmath}
\usepackage{amsthm}
\usepackage{mathtools, bm, nccmath}
\usepackage{braket}
\usepackage{amsfonts} 
\usepackage{amssymb}
\usepackage{calrsfs}
\usepackage{ bbold }
\usepackage{ tipa }
\usepackage{enumitem}
\usepackage{eucal}
\usepackage{cancel}
\usepackage{algorithm}
\usepackage[noend]{algpseudocode}
\usepackage{ wasysym }
\usepackage{upgreek}
\usepackage{nicefrac}
\usepackage{caption}
\usepackage[square,numbers,sort&compress]{natbib}
\usepackage{float}
\usepackage[section]{placeins}
\usepackage{xcolor}
\usepackage[caption=false]{subfig}

\newcommand{\average}[1]{\langle #1 \rangle}

\newtheorem{prop}{Proposition}

\newtheorem{definition}{Definition}[]
\newtheorem*{remark}{Remark}

\begin{document}

\title{Distributing Multipartite Entanglement over Noisy Quantum Networks }

\author{Lu\'{i}s Bugalho}
\affiliation{Instituto Superior Técnico, Universidade de Lisboa, Portugal}
\affiliation{Physics of Information and Quantum Technologies Group, Centro de Física e Engenharia de Materiais Avançados (CeFEMA), Portugal}
\affiliation{PQI -- Portuguese Quantum Institute, Portugal}
\email{luis.bugalho@tecnico.ulisboa.pt}
\author{Bruno C. Coutinho}%
\affiliation{Instituto de Telecomunica\c{c}\~{o}es, Portugal}
\author{Francisco A. Monteiro}
\affiliation{Instituto de Telecomunica\c{c}\~{o}es, Portugal}
\affiliation{ISCTE - Instituto Universitário de Lisboa, Portugal}
\author{Yasser Omar}
\affiliation{Instituto Superior Técnico, Universidade de Lisboa, Portugal}
\affiliation{Physics of Information and Quantum Technologies Group, Centro de Física e Engenharia de Materiais Avançados (CeFEMA), Portugal}
\affiliation{PQI -- Portuguese Quantum Institute, Portugal}

\maketitle

\begin{abstract}
A quantum internet aims at harnessing networked quantum technologies, namely by distributing bipartite entanglement between distant nodes. However, multipartite entanglement between the nodes may empower the quantum internet for additional or better applications for communications, sensing, and computation. In this work, we present an algorithm for generating multipartite entanglement between different nodes of a quantum network with noisy quantum repeaters and imperfect quantum memories, where the links are entangled pairs. Our algorithm is optimal for GHZ states with 3 qubits, maximising simultaneously the final state fidelity and the rate of entanglement distribution. Furthermore, we determine the conditions yielding this simultaneous optimality for GHZ states with a higher number of qubits, and for other types of multipartite entanglement. Our algorithm is general also in the sense that it can optimize simultaneously arbitrary parameters. This work opens the way to optimally generate multipartite quantum correlations over noisy quantum networks, an important resource for distributed quantum technologies.
\end{abstract}
%\tableofcontents

\section{Introduction}

Quantum technologies hold the promise of faster computing, securer private communications \cite{Bennett2014,Nurhadi2018,Broadbent2009}, and more precise sensing and metrology \cite{Chuang2000,Gottesman2012}. Quantum networks open the possibility to explore these applications in distributed scenarios, allowing for increased performance and/or tasks involving multiple parties. In fact, quantum networks where the links correspond to quantum entanglement between the nodes, can be thought over long distances, \textit{e.g.} inter-city, such as in the case of a quantum Internet \cite{Wehner2018,Pompili2021a}, or locally, \textit{e.g.} inside a laboratory or a quantum local-area network (QLAN) \cite{Alshowkan2021a}.

The immediate question that arises on a quantum network is how to optimally generate bipartite entanglement between two user-nodes, as a function of the relevant metrics (fidelity, rate,...). This encompasses selecting the appropriate protocols for entanglement generation and entanglement swapping \cite{Munro2015}, and algorithms that find the optimal way to execute them over a quantum network \cite{Caleffi2017,Chakraborty2019a,Shi2019b,Li2020,Dai2020,Bauml2020,Santos2023}.

However, a number of applications go beyond the two-party paradigm and require multipartite entanglement, a framework with no classical analogue. Important examples of these applications are quantum sensor networks \cite{Ren2012,Khabiboulline2019,Eldredge2018,Qian2020}, multi-party quantum communication \cite{Hillery1999,Zhu2015,Murta2020} and distributed quantum computation \cite{DHondt2004,Raussendorf2001}. For the distribution of multipartite entanglement, theoretical upper bounds derived from the communication capacities have been studied \cite{Laurenza2017,Pirandola2019,Pirandola2019a,Pirandola2020,Das2021a,Bauml2020} and several distribution schemes have already been developed \cite{Meignant2018,Wallnofer2019,Goodenough2020}. 

In this work we aim at finding the optimal way to distribute multipartite entanglement in noisy quantum networks, under a given distribution scheme. This has particular relevance for applications where noise and the distribution of the state \cite{Filippov2013} impacts the application itself. To that end, we introduce a new methodology that allows to maximize two different objectives -- the rate of distribution and the fidelity of the distributed state -- even though our approach is easily generalizable to include more. We develop an algorithm with tools from classical routing theory \cite{Sobrinho2005,Demeyer2013a} that finds the optimal way of distributing a 3-qubit GHZ state, providing that the metrics that describe its distribution follow a set of properties, which we determine. We also find the conditions that yield optimality of our algorithm when considering a higher number of qubits, under one of the possible schemes. Moreover, our methodology is adaptable to different underlying physical implementations of the constituents of a quantum internet and its different stages of development \cite{Wehner2018}.

\section{Quantum Network Description \label{section:network}}

Let us first describe our model of a quantum network. Structurally, it is characterised by a graph $G(V,E)$ where each node is denoted by a letter $j \in V$ and the link connecting nodes $i$ and $j$ is denoted by $i:j \in E$. Each path is a sequence of links and we identify them by their initial and final node $m:n$.  Note that even though we only use the first and last node to represent a path, it is just a matter of notation, as paths are concatenations of contiguous links, and there usually exist different paths between any two nodes.  The set of nodes $\mathcal{T} \subset V$ we want to distribute the state to is called the terminal set, where $|\mathcal{T}| = T$ is the number of terminal nodes, or equivalently, the number of qubits of the distributed state. Each link is a noisy quantum channel, accompanied by a classical channel for signaling success and sending corrections, connecting neighboring nodes which individually hold qubits in imperfect quantum memories . Each quantum channel is capable of establishing entanglement between its two nodes and each quantum node must be able to realize one and two-qubits unitary gates, as well as performing measurements on its qubits \color{black} \cite{Wehner2018,Brand2020,Goodenough2020,Meignant2018} \color{black}. The protocols for entanglement generation and entanglement swapping are considered to be probabilistic, following a geometric distribution in the number of trials before the first success \cite{Brand2020}. Moreover, each entangled pair of the network is modelled by a Werner state \cite{Werner1989} $\rho_W = \gamma  \ket{\phi^+} \bra{\phi^+} + (1-\gamma)/4 \ \mathbb{1}$, where $\ket{\phi^+} \propto \ket{00} + \ket{11}$ is a maximally entangled state, and with $\gamma = (4F-1)/3$, where $F$ is the fidelity of the state. In the bipartite case, there is an equivalence between a Werner state and a representation with a depolarising channel having an equal amount of bit-flip, phase-flip and phase-bit-flip errors: $\rho_W = \mathcal{D}_1^F (\ket{\phi^+} \bra{\phi^+}) = \mathcal{D}_2^F (\ket{\phi^+} \bra{\phi^+})$. This quantum channel is described in two alternative forms, one of them using the partial transposition operator $\Lambda_i$ on qubit $i$:

\begin{equation}
\begin{aligned}
	\mathcal{D}_i^p (\rho) =&\ p \rho + \frac{1-p}{3}(\hat{X}_i \rho \hat{X}_i^\dagger + \hat{Y}_i \rho \hat{Y}_i^\dagger + \hat{Z}_i \rho \hat{Z}_i^\dagger) \\
	=&\ \frac{1+2p}{3} \rho + \frac{2(1-p)}{3} \Lambda_i (\hat{Y}_i \rho \hat{Y}_i^\dagger) .
	\label{eq:DepolarisingChannel}
\end{aligned}
\end{equation}
where the parameter $p$ controls the amount of error in the state and $\hat{X}_i, \hat{Y}_i, \hat{Z}_i$ are the usual Pauli gates acting on qubit $i$. This equivalence is significant to find the final form of the distributed state. In addition, this trace-preserving map has important properties, among them linearity on the main argument $\rho$, commutativity for every index $i$, and invariance under unitary operations applied to single qubits \cite{Hein2006}.

\section{Distribution of Entanglement}

Distributing entanglement is the building block for an entanglement-based quantum network. There are various schemes for doing such, whether be it in a bipartite scenario, between two parties, usually in the form of a Bell pair, or in the multipartite case, whereas there are more than two parties and multiple possible states, not all equivalent nor with identical applications. We start by first describing the case for bipartite entanglement, as it is often the first step of possible schemes for the multipartite case, for example the star scheme, which we detail next for the case of a 3-Qubit GHZ state.  From \cite{Meignant2018}, we make use of the protocol for distributing this state, that minimizes the number of consumed entangled pairs, and adapt it to our multi-objective optimization.  In Appendices \ref{appendix:treescheme} and \ref{appendix:wstate} we extend these results for a different scheme of distribution \cite{Meignant2018} and W states, respectively.

In both cases we identify metrics that attribute meaningful weights to a certain property of the distribution across the network. We do this in a way that guarantees each metric follows a set of properties (detailed in a Section \ref{section:algorithms}), so to ensure that our algorithms converge to the optimal solution.

\subsection{Distribution of Bipartite Entanglement \label{section:bipartite}}

To describe the metrics involved in distributing Bell pairs across a chain of quantum links, $i.e.$ a path, we must concern the protocols to do so. These protocols are those of entanglement generation between neighboring nodes and entanglement swapping to extend the range of the entanglement. They will impact four chosen properties of the network: \textit{(i)} the probability of success and \textit{(ii)} communications times, which are self-explanatory, the \textit{(iii)} fidelity of the state, a measure of how much noise exists in a quantum state, and finally the \textit{(iv)} memory coherence time, a characteristic of a quantum memory that measures how much time can a qubit be in the memory before the state is useless due to noise.  Note that the individual protocols for the entanglement generation in each of the links can be considered as a separate problem. The values for the probability of success, for the communication times and for the fidelity for each link are results from the entanglement generation protocols used. This makes each link agnostic from the physical implementation of the protocol. Nonetheless, we require a characterization of the protocol in these three quantities which are the input for the algorithms. 

Starting with the probability of success, denote by $p_{i:j}$ the probability of successful entanglement generation between neighboring nodes $i$ and $j$ at the first attempt and by $k_{j}$ the probability of successful entanglement swapping on node $j$ also at the first attempt. Considering every stochastic process to be independent, the probability of success of generating end-to-end entanglement across a chain of quantum nodes will also be geometrically distributed with a probability of success at the first attempt given by:

\begin{equation}
    p_{m:n} = \prod_{i:j \in m:n} p_{i:j} \prod_{j \in m:n \setminus m,n} k_j.
\end{equation}

The classical communications time will also play an important role in the rate and each branch fidelity. We denote by $t_{i:j} = L_{i:j}/c$ the time it takes to send a classical message between neighboring nodes $i$ and $j$, distanced by $L_{i:j}$. In our simplified model, there are two rounds of classical communication across the chain: \textit{i)} one for communicating successful entanglement generation across every link of the chain, and \textit{ii)} another to communicate successful entanglement swapping in every node of the chain, with correspondent corrections. This takes two times the sum of each individual time across the chain to complete each successful round:

\begin{equation}
    t_{m:n} =  2 \sum_{i:j \in m:n} t_{i:j}.    
\end{equation}

As for the fidelity across a chain after the entanglement swapping was completed, taking advantage of the $\gamma$ change of variables, for Werner states,  described in Section \ref{section:network} where $\gamma = (4F-1)/3$ \cite{Werner1989} , results in a simple multiplication of each $\gamma$ value along the chain.  Note that this fidelity can also be affected by noisy operations required for performing entanglement swapping. In this case, one can include the errors from these operations directly in the fidelity of the entangled pairs via a mathematical trick \cite{Dur2007}, reducing this problem to simply a new value for the fidelity of each link as before.  When using $\gamma$ we have a threshold value at $1/3$ (correspondent to $F=1/2$), meaning that below this threshold, entanglement does no longer exist and the path cannot be used:

\begin{equation}
    \gamma_{m:n} = \begin{cases}
        \underset{i:j \in m:n}{\prod} \gamma_{i:j} \ ,& \ \gamma_{m:n} \geq 1/3 \\
        0 \ ,& \  \gamma_{m:n} < 1/3.
    \end{cases}
\end{equation}

In order to include the quantum memory coherence time, similarly to what has been done in \cite{Guo2017,Brand2020}, we adapted the contributions to our simpler scheme. Consider that, for Bell states, the effect of memory decoherence in the fidelity of the state verifies $\gamma \mapsto \gamma  e^{- t_{\text{wait}}/\sigma}$, where $\sigma$ is the quantum memory coherence time. In the first round, if the generation is successful, entanglement is stored between every neighboring node of the chain. Therefore, the memory decoherence factor of each node, apart from the first and the last, contributes twice to the final state distribution. On the second round, if the swapping is successful, the entanglement is held only within the first and the last node. This results in every node contributing the same to the decoherence factor, given that the communications time, $t_{wait}$, is identical:

\begin{align}
	\frac{1}{\sigma_{m:n}} = \sum_{i \in m:n} \frac{2}{\sigma_i}.
\end{align}

\begin{figure}[t]
\includegraphics[width=\linewidth]{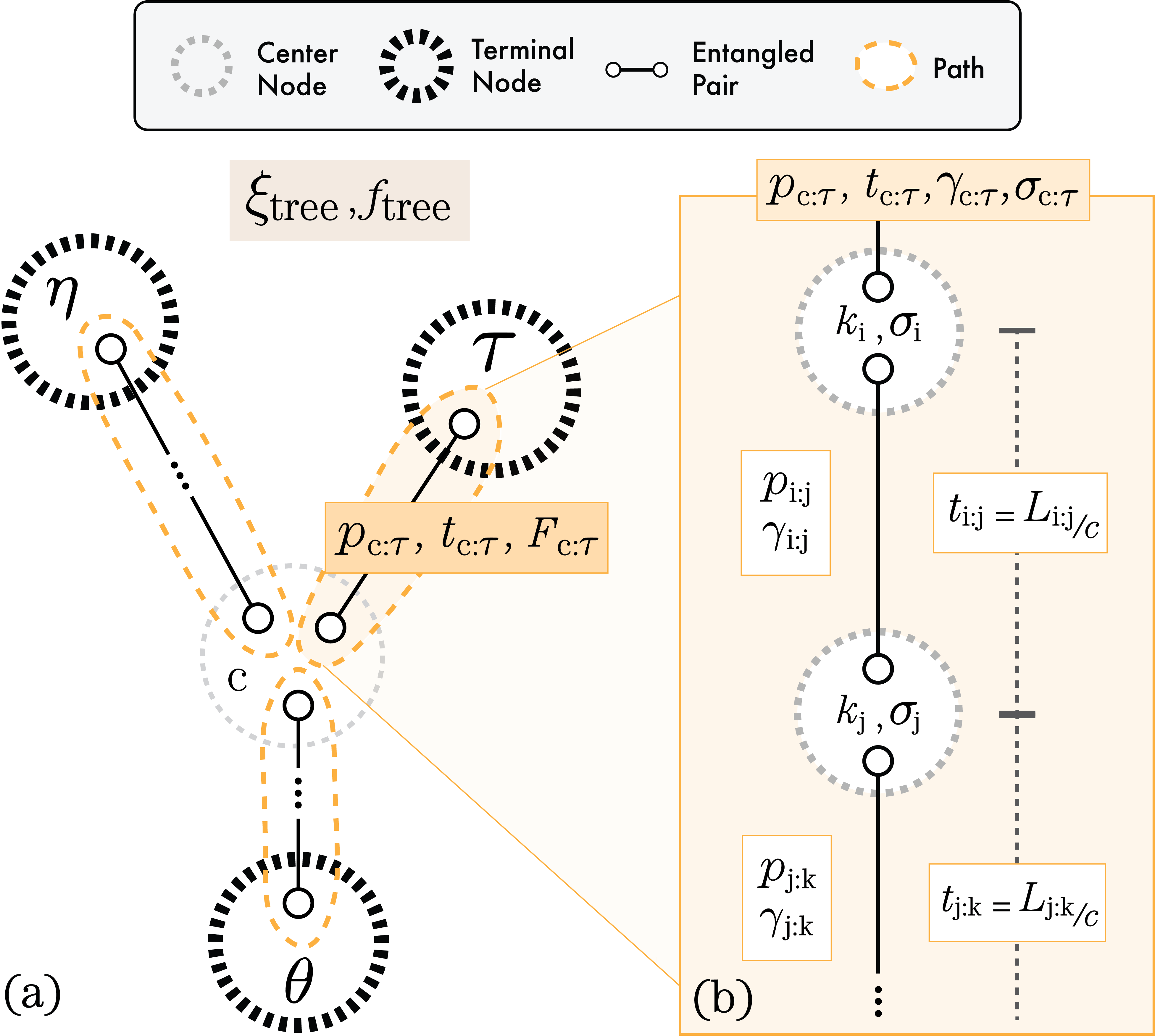}
\caption{ (\emph{a}) Star composed by 3 different paths, connecting terminal nodes $\tau$, $\eta$ and $\theta$ to the center node $c$. Notice that the rate $\xi_{tree}$ depends on each branch probability of success $p_{c:\tau}$ and communications time $t_{c:\tau}$ (\emph{Eq. \ref{eq:rate}}). Moreover, the fidelity of the final state $f_{tree}$ depends on each branch fidelity $F_{c:\tau}$ (\emph{Eq. \ref{eq:fidelity}}). (\emph{b}) Branch composed by an arbitrary number of links. Each branch final fidelity $F_{c:\tau}$ depends on the path values $\gamma_{i:j}$, communications time $t_{i:j}$, and memory decoherence times $\sigma_{i:j}$. The probability of success $p_{c:\tau}$ depends on each link probability $p_{i:j}$ and the entanglement swapping probabilities $k_j$. The communications time $t_{c:\tau}$ depends on the lengths and the velocity of light $L_{i:j}/c$. }
\label{fig:scheme3}
\end{figure}

These four parameters can then be contracted to only three:

\begin{align}
        [p_{m:n},t_{m:n},\gamma_{m:n},\sigma_{m:n}] &\mapsto [p_{m:n},t_{m:n},F_{m:n}] , 
        \label{eq:fidelitydecoherence}
\end{align}
subject to $(4F_{m:n}-1)/3 = \gamma_{m:n} e^{-t_{m:n}/\sigma_{m:n}}$.  The final parameters in \emph{Eq. \ref{eq:fidelitydecoherence}} are then the input of the multipartite metrics that characterize the fidelity and rate of the final multipartite state. In \emph{Fig.\ref{fig:scheme3}} we present a scheme of each of the detailed parameters, and how they affect the final metrics.  Nonetheless, the four parameters that are the input of \emph{Eq. \ref{eq:fidelitydecoherence}} must be routed separately for the optimal paths (and consequently the optimal star) to be found, under appropriate algorithms.

\subsection{Distribution of 3-Qubit GHZ states} 

Moving on to the multipartite case, let us consider the distribution of a GHZ state with three qubits. Similarly to what has been done in \cite{Meignant2018}, to distribute GHZ states one would have to first find the Steiner tree connecting the set of terminal nodes and then perform some measurements over the Steiner nodes.  The so called Steiner tree is the shortest-tree connecting the terminal nodes, where a tree is a graph with no cycles or, equivalently, only one path connecting any two nodes.  We start with the case of 3 qubits, and later generalise for an higher number of qubits. The reason is that the Steiner tree connecting 3 terminal nodes is always a star-graph making the scheme equivalent to first distributing bipartite entanglement between every terminal and a center node and then performing a set of operations on the center node. Thus, the central idea is finding the center node for which this results in the optimal solution. To define optimality, we again concern the metrics for the distribution of this entangled state, namely the rate of distribution and the final state fidelity.

Consider that the center node, $c$, is capable of coordinating every process. There is an initial phase to signal the beginning of bipartite entanglement distribution, followed by the phase of creating bipartite entanglement in every branch connecting to the center node and finally performing the set of operations that result in the desired state distribution. Following \cite{Caleffi2017}, we can provide a measure for the average rate of distribution, $\xi_{tree}$, considering the different steps in this scheme. This average rate is given by the inverse of the average time, $t_{tree}$, it takes to complete the distribution. Considering a scheme of attempts until the first success, if any branch fails to create bipartite entanglement, every branch starts over, the final distribution of the probability of the first success will follow a geometric distribution.
In such case, the average time it takes to completely distribute the multipartite entangled state is the communications time multiplied by the average number of attempts before the first success:
\begin{equation}
\begin{aligned}
	t_{\text{tree}} &=  \frac{ 2\cdot \max_{\tau \in \mathcal{T}} \{ t_{c:\tau} \}  }{\prod_{\tau \in \mathcal{T}} p_{c:\tau}}  \quad  \\ 
	\xi_{\text{tree}} &= \frac{1}{t_{\text{tree}}}
	\label{eq:rate}   
\end{aligned},
\end{equation}
where $p_{c:\tau}$ and $t_{c:\tau}$ correspond, respectively, to the probability of success and communications time of the path $c:\tau$. If the operations performed at each Steiner node are not deterministic, this time will increase by multiplying the expected value of the stochastic process that characterizes the global set of operations.  Moreover, one could also pursue an approach similar to the one in \cite{Kamin2022a} to derive an average rate for the case where the center node waits for each path to conclude generating end-to-end entanglement.  

The fidelity can be calculated by applying a depolarising channel to each terminal node substituting the $p$ value in \emph{Eq. \ref{eq:DepolarisingChannel}} by the fidelity  $F_{c:\tau} \equiv F_{\tau}$, of each branch connecting the center node $c$ to the terminal $\tau$. Using the second description of the depolarising channel, the result is:

\begin{equation}
\begin{aligned}
	f_{\text{tree}} = \frac{1}{2} \Bigg[ \prod_{\tau \in \mathcal{T} } \frac{1+2F_{\tau}}{3} +  \prod_{\tau \in \mathcal{T} } \frac{2(1-F_{\tau})}{3} +& \\
	+ \prod_{\tau \in \mathcal{T}} \frac{4F_{\tau} -1}{3}&  \Bigg]
	\label{eq:fidelity}
\end{aligned}
\end{equation}

Under \emph{Eqs. \ref{eq:rate}-\ref{eq:fidelity}}, we can map the $|\mathcal{T}|$ paths' parameters to the tree's rate of distribution, $\xi_{tree}$, and fidelity of the final state, $f_{tree}$: 

\begin{equation}
    \begin{aligned}
        \Big\{ [p_{c:\tau},t_{c:\tau},F_{c:\tau}]\Big\}_{\tau \in \mathcal{T}} &\mapsto [\xi_{\text{tree}},f_{\text{tree}}]
        \label{eq:mappathstree}
    \end{aligned}
\end{equation}

 These metrics output the values for the fidelity of the distributed state and rate of the distribution. They depend necessarily on the network, and on the parameters of the individual links and nodes, which are a result of the physical implementation. Therefore, the fidelity and the rate at which one can distribute  GHZ states across the network depends on the underlying physical implementation, the chosen distribution protocols and parameters of noise of the setup. Nonetheless, as will be seen ahead, the routing algorithm only takes the parameters, the metrics, and outputs the optimal solutions for the distribution. 

\section{Algorithms for Optimal Multipartite Entanglement Distribution \label{section:algorithms}} 

Now that we can characterize both the rate of distribution and fidelity of the final state, we need to find the optimal star. Here, the definition of optimal is crucial. In the case we only deal with one parameter, $e.g.$ the fidelity, there only exists one optimal solution, unless several solutions have the same fidelity. However, if we want to consider more parameters to optimize, a multi-objective approach is necessary. In \cite{Martins1984}, multi-objective shortest-paths (MOSP) algorithms are introduced by defining the dominance relation and using the Pareto optimality definition, as we will explain further ahead. This results in a set of optimal solutions that is usually larger than one -- the Pareto front. The same relation is also valid and central for finding the optimal way to distribute a 3-qubit GHZ state. In \cite{Sobrinho2003,Sobrinho2005}, a more fundamental definition for these metrics for routing problems is detailed using algebras for routing, which we use inadvertently to refer to metrics as well. 

\theoremstyle{definition}
\begin{definition}\emph{Algebra for Routing} is an ordered septet $(W,\preceq,L,\Sigma,\phi,\oplus,f)$ comprised as follows: $W$ a set of weights, $\preceq$ a total order, $L$ a set of labels, $\Sigma$ a set of signatures, $\phi$ a special signature, $\oplus$ a binary operation that maps pairs of labels and signatures into a signature, and a function $f$ that maps signatures into weights.
\end{definition}

\begin{figure}[t]
\centering
\includegraphics[width=\linewidth]{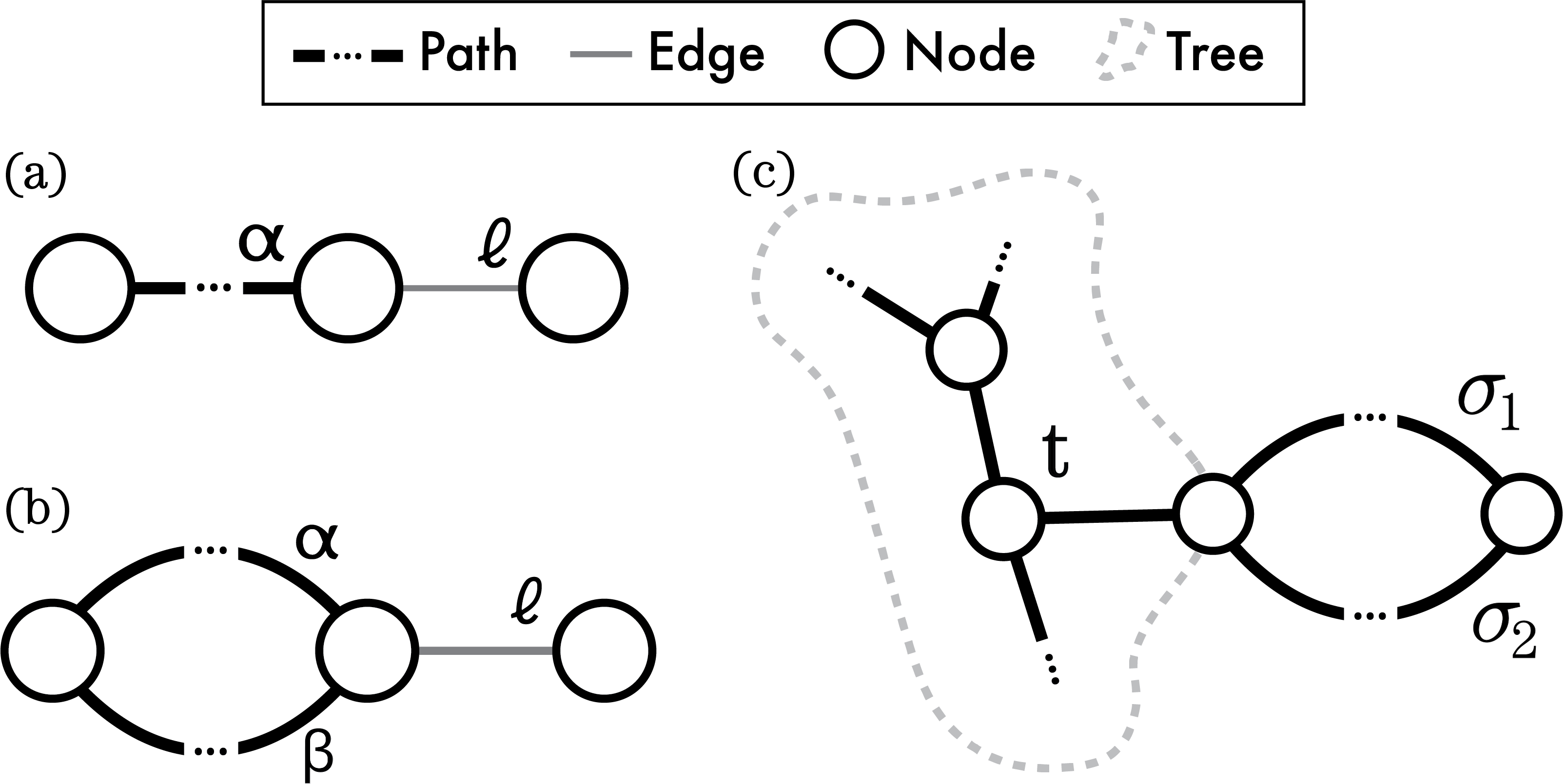} 
\caption{Illustration of the context in the algebras' properties as found on the definitions of: (\textit{a}) \textbf{Monotonicity}: extending a path $\alpha$ with a link $l$, results in a worse path; (\textit{b}) \textbf{Isotonicity}: if path $\alpha$ is better than path $\beta$, then any extension with any link $l$, maintains the order; (\textit{c}) \textbf{Label-isotonicity}: if path $\sigma_1$ is better than path $\sigma_2$, then any tree $t$ extended by path $\sigma_1$ is better than extending with path $\sigma_2$. }
\label{fig:properties}
\end{figure}

 Intuitively, these algebras are a mathematical object that transports each important aspect of routing, onto a mathematical object related with each concept. Namely: the weight paths can have belong to $W$, to each link corresponds a label in $L$, to each path corresponds a signature in $\Sigma$, binary operations $\oplus$ are used to define how paths are extended and total-orders $\preceq$ to define the ordering of the parameters, $i.e.$, which one is the better. There is also a special signature $\phi$ correspondent to the impossible path, important for disregarding some paths.  Using this definition, some properties can be defined \cite{Sobrinho2005}, guaranteeing optimality when finding the shortest-path with an appropriate algorithm (for intuition refer to Fig. \ref{fig:properties}):

\theoremstyle{definition}
\begin{definition}{(Monotonicity)}
an algebra for routing is called monotone if: $\forall \ l \in L; \alpha \in \Sigma: f(\alpha) \preceq f(\alpha \oplus l)$.
\label{def:monotonicity}
\end{definition}

\theoremstyle{definition}
\begin{definition}{(Isotonicity)}
an algebra for routing is called isotone if: $
\forall  \ l \in L; \alpha,\beta \in \Sigma : f(\alpha) \preceq f(\beta) \Rightarrow f(\alpha \oplus l) \preceq f(\beta \oplus l)$.
\label{def:isotonicity}
\end{definition}

Given the multi-objective scenario analyzed, the dominance relation becomes essential. This comes from the fact that, while one path might be better for some parameters, another path might be better for other parameters.  Given a set of algebras for routing $\{ (W^i, \preceq^i , L^i ,  \Sigma^i , \phi^i , \oplus^i ,  f^i )\}$, each one characterizes one of the objectives and associated set of parameters, where $i \in \{1,...,k\}$, being $k$ the total number of objectives, or the total number of algebras.  The dominance relation is then given by the following definition, as detailed in \cite{Martins1984}:

\theoremstyle{definition}
\begin{definition}{(Dominance)\label{def:dominance}}
let $\omega$ and $\nu$ be two different signatures in $\{ \Sigma^i\}$. We call the relation $D$ dominance, and we say $\omega$ dominates $\nu \equiv \omega \  \text{D} \ \nu$, if $f^j(\omega^j) \preceq^j f^j(\nu^j)  \ \forall j \in \{ 1, ..., k \}$ and the strict order holds at least once. 
\end{definition}

This means that one path only dominates the other if it is better in every single aspect.  This causes some paths not to dominate another, nor be dominated by other, $e.g.$, path $a$ has a better rate than path $b$, but path $b$ has a better fidelity than path $a$.  Unlike, for example, greater or equal than, this is not a total order , where the total means that for every two weights $f(a)$ and $f(b)$ we have that either $f(a) \preceq f(b)$ or $f(b) \preceq f(a)$. From the example above, we can verify that sometimes path $a$ does not dominate path $b$, and path $b$ does not dominate path $a$ either.  This is the fundamental reason why we could not create an algebra for routing with the dominance relation defining the order of the algebra.

In the same manner, an algebra for trees can also be created. In this case, the labels define paths, rather than of edges ($L \rightarrow \Sigma$), and signatures define trees, rather than of paths ($\Sigma \rightarrow \Xi$). Using these algebras for trees, another property arises, enabling that the shortest-tree necessarily contains the possible shortest-paths if the schemes for creating paths and trees are different. This property is what we defined as label-isotonicity:

\theoremstyle{definition}
\begin{definition}{(Label-isotonicity)}
An algebra for trees $(W,\preceq,\Sigma,\Xi,\phi,\oplus,f)$ is said to be label-isotone if \footnote{By $\sigma_1 \preceq \sigma_2$ we mean $f(\sigma_1 \oplus e_{\Sigma}) \preceq f(\sigma_2 \oplus e_{\Sigma})$ where $e_\Sigma$ is the neutral element of $\Sigma$ w.r.t $\oplus$.}: $\forall  \ \sigma_1, \sigma_2  \in \Sigma; t \in \Xi : \sigma_1 \preceq \sigma_2 \Rightarrow f (t \oplus \sigma_1) \preceq f(t \oplus \sigma_2) $
\label{def:pathisotonicity}
\end{definition}

From the scheme for creating a 3-qubit GHZ state, the main problem is finding the correct center node, such that the corresponding star is optimal.  Thus, one first needs to find the shortest-paths between each of the terminal nodes and all the other nodes. This requires running the MOSP routine $T$ times, followed by adding non-dominated trees to the set of solutions while attempting every combinations of paths for each center-node. Moreover, this center node must be reachable from each of the terminals, allowing the search to be optimized:

\begin{algorithm}[H]
\caption{Algorithm for T-Star} \label{alg:starexact} 
\begin{algorithmic}[1]
	\Procedure{T-Star Exact}{$terminal$} 
		\State $A :=$ Set of solutions initialized empty;
		\For{$node \in terminal$}
			\Procedure{Shortest-path}{$node$}; \EndProcedure	
		\EndFor
		\State $Nodes_{reach} \gets$ Nodes reachable from every terminal;
		\For{$node \in Nodes_{reach}$}
			\For{Tree T = $\cup_{i,j} Path^j(node,terminal_i)$}
				\If{T is non-dominated by every tree in A}
					\State Add T to A;
				\EndIf
			\EndFor
		\EndFor
	\EndProcedure
\end{algorithmic}
\end{algorithm}

{
\renewcommand{\thetheorem}{\ref{th:treeshortest}}
\begin{prop}
For the shortest-star with $3$ terminals, the paths connecting the center node to the terminals must be the shortest-paths, if the underlying algebras for trees are label-isotone.
\end{prop}
\addtocounter{prop}{-1}
}

\emph{Proposition \ref{th:treeshortest}} provides that \emph{Algorithm \ref{alg:starexact}} converges to the set of optimal solutions for the 3-qubit GHZ state problem. Considering that the solutions for the MOSP algorithms are optimal (see Appendix \ref{appendix:bipartite}), it can also be verified that the rate of distribution and fidelity of the final state are both monotone and label-isotone for any number of qubits in the star scheme. This ensures the optimality of \emph{Algorithm \ref{alg:starexact}}. We present the proof the algorithm converges to the optimal solution under the algebra properties in Appendix \ref{appendix:proof} and that all the multipartite metrics for GHZ and W states verify these properties in Appendices \ref{appendix:monotonicity}, \ref{appendix:label} and \ref{appendix:Wproperties}.

\subsection{Distribution of N-Qubit GHZ states}

When considering a higher number of qubits, there are two possible schemes to distribute this state. The first is the \textit{star scheme}, which is an extension of the one introduced resembling a star-graph, hence its name. Identically, it consists of generating bipartite entanglement between every terminal node and a center node, which can ultimately be any of the terminal nodes, and then projecting every qubit of the center node in the desired state. This scheme is capable of distributing any multipartite entangled state, besides GHZ states, for example W states, which we also detail and prove optimality in Appendices \ref{appendix:wstate} and \ref{appendix:Wproperties}. Using \emph{Algorithm \ref{alg:starexact}} also leads to optimality under the same constraints on the algebras, if the same link can be used more than once. If the latter is untrue and this condition is only imposed $a\ posteriori$, $i.e.$, from the set of solutions from the algorithm, taking only the ones where each link is used only once, then we can only extract a part of the set of optimal solutions and flag if there might exist other optimal solutions or not. Note that for three qubits, this condition did not matter since, if there was an overlap of links used, then there is always another better star, with the center node located elsewhere. 

When we do not restrict ourselves to finding the best star, but instead try to find the best tree, then we should use the \textit{tree scheme} \cite{Meignant2018}, detailed in Appendix \ref{appendix:treescheme}, alongside with a multi-objective Steiner tree algorithm. Although this is always either equal or better at distributing GHZ states, the algorithm is much more complex. It is more efficient when the number of nodes grows larger, since it almost surely reduces the number of entangled pairs consumed. However, it cannot be used to distribute some states, such as a W state. 

\subsection{Simulations of the Algorithm}

\begin{figure}[t]
\centering
\includegraphics[width=\linewidth]{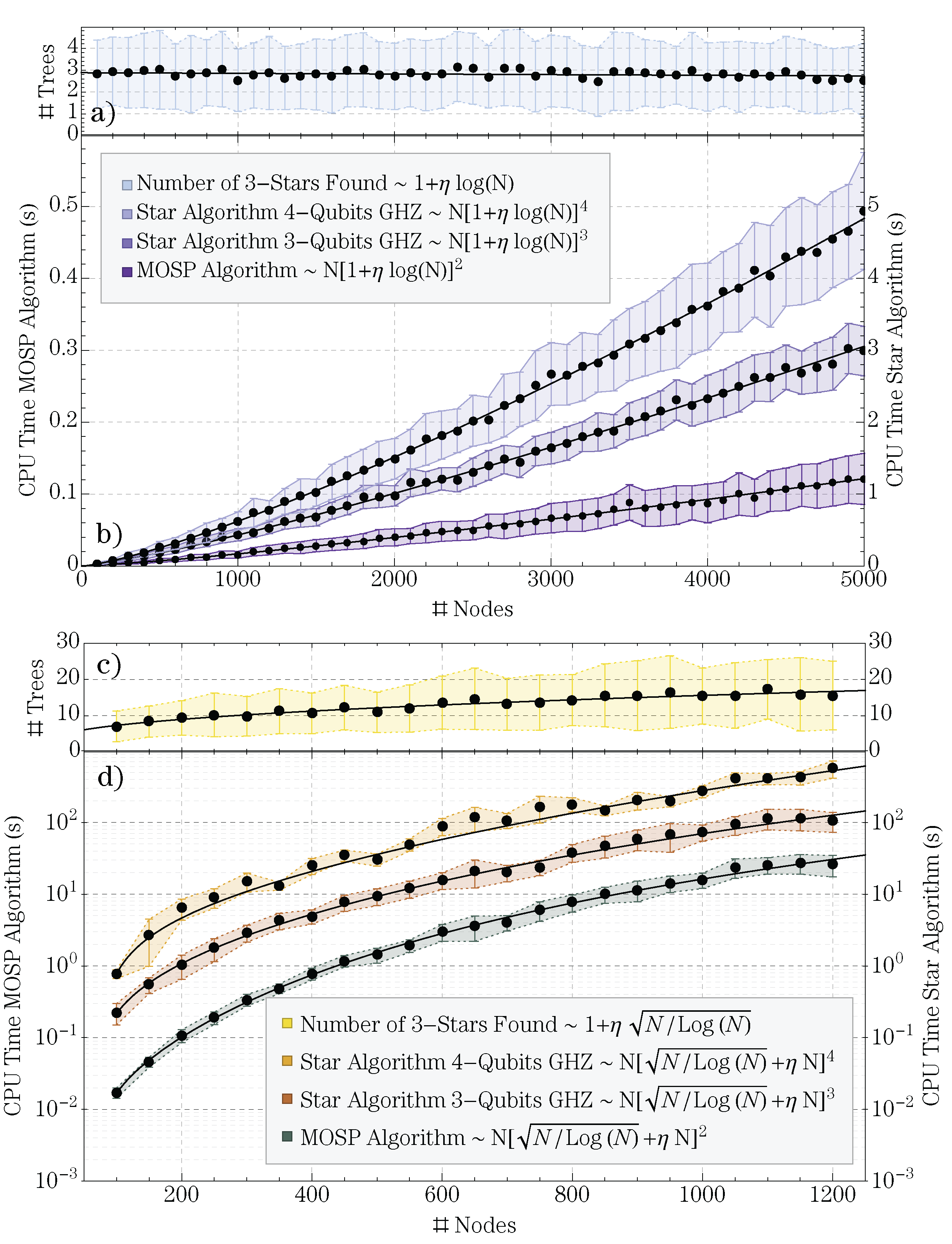}
\caption{Simulations for MOSP algorithm and \emph{Algorithm \ref{alg:starexact}} for 3 and 4 qubits GHZ state distribution in (\emph{b}) Erdös-Rényi networks with average degree $\average{\lambda} = 3$  and (\emph{d}) random geometric networks with average degree $\average{\lambda} = 8$, displaying as well the average number of solutions found for \emph{Algorithm \ref{alg:starexact}} applied to the 3-qubits problem in (\emph{a}) Erdös-Rényi networks and (\emph{c}) random geometric networks. The parameters utilized are: $p_\text{min} = 0.5$, $t_\text{min}=1$, $t_\text{max}=100$,  $\sigma_\text{min} = 10^4$, $\sigma_\text{max} =10^5$, $f_\text{trunc}^\text{GHZ}=0.5$, $f_\text{trunc}^\text{path} = 0.9$. For a Erdös-Rényi network $a = 2$ and $d_\text{max} = \log{N}/\log{\average{\lambda}}$ \cite{Barabasi2016,Dorogovtsev2008}. For random geometric networks $a = 2$ and $d_\text{max} = \sqrt{N/\log{N}}$ \cite{Ellis2007}. } 
\label{fig:TreeTerminals}
\end{figure} 

In \emph{Figure \ref{fig:TreeTerminals}} we present the scaling of the complexity of this algorithm in two different types of networks, namely Erdös-Rényi networks \cite{Barabasi2016,Dorogovtsev2008}, which capture some properties of a quantum internet, namely the small-world property, and random geometric networks \cite{Dall2002}, which convey the limitation in the length of each individual link in a quantum network \cite{Inagaki2013}.

Erdös-Rényi networks \cite{Barabasi2016,Dorogovtsev2008} are networks where the degree of a node, $i.e.$, the number of neighbors a node has, follows a Poisson distribution characterized by an average degree $\average{\lambda}$. To build this network, one creates all the nodes and decides if a link between two nodes is attributed based on a probability $p$. One can recover the limit of Poisson distribution when the number of nodes is large enough $n \rightarrow \infty$ and $np \rightarrow const = \average{\lambda}$: $\operatorname{Pr}(degree(v)=k)=\average{\lambda}^{k} e^{-\average{\lambda}}/k!$.

On the other hand, random geometric networks establish links if two nodes are close enough, a property a quantum internet verifies locally. To create a random geometric network one generates a grid of equal sides and generates $n$ nodes with randomly distributed $xy$ positions inside the grid. Then, depending on a graph property of the graph, the radius $r$, if two of these nodes are closer to each other than the radius, a link is established between such nodes. This radius $r$ is intimately connected with the average degree of the graph \cite{Dall2002}.

Every link has a value of fidelity uniformly distributed in $[f_{\text{min}},1)$, and a probability of successful entanglement generation and probability of successful entanglement swapping distributed uniformly in $[p_{\text{min}},1)$. The communications time and memory coherence time are distributed uniformly in $[t_\text{min},t_\text{max}]$ and $[\sigma_\text{min},\sigma_\text{max}]$, respectively. The values of $f_\text{min}$ scale approximately through a power law $f_\text{min} \sim [f_\text{trunc}]^{a / d_\text{max}}$ \cite{Coutinho2022} that guarantee each path contains entanglement and the final distributed state fidelity $f_\text{trunc}^\text{GHZ} > 1/2$.  In the previous expression, we report $d_\text{max}$ to the diameter of a graph, which is the maximum distance of the shortest-path between any two nodes of the graph. This scaling is a way to partially guarantee the functional connectivity of the network, $i.e.$, that one can find a path with a fidelity above 1/2 between any two points of the network. 

Usually, the minimum value for the fidelity of a path $f_\text{trunc}$ is $1/2$, for which entanglement is known to be present. However, in our case, to guarantee that most of the simulations would result in finding at least one optimal stars with $f_\text{trunc}^\text{GHZ} > 1/2$, we set the value for each path $f_\text{trunc} = 0.9$. 
Using these values and all the metrics described throughout the paper, one can verify the properties necessary to prove the algorithm will find the optimal solutions (see Appendix \ref{appendix:bipartite} for the bipartite case and Appendices \ref{appendix:monotonicity}, \ref{appendix:label} and \ref{appendix:Wproperties} for the multipartite extension).

\color{black}
\subsection{Comparison with existing Multipartite Routing Algorithms}

To compare with previous results presented in the literature, we made an analysis for two different routing protocols to distribute GHZ states with three qubits. The first method was based on Ref. \cite{Meignant2018}, where performing routing to optimize the distribution of a GHZ state is identical to finding the shortest-star minimizing the communications time and the amount of entangled pairs consumed (or links).  We call this a \textit{shortest-path} routing method. This derives from the fact that the pairs are considered to be pure states, the probability of success is one, and all links have the same length. This case is the most intuitive one, as the communications time is bounded by the maximum length of each path connecting to the a terminal, and the lesser the number of consumed pairs, the higher are the values for the rate and fidelity that could potentially be achieved.
\begin{figure}[t]
    \centering
    \includegraphics[width=\columnwidth]{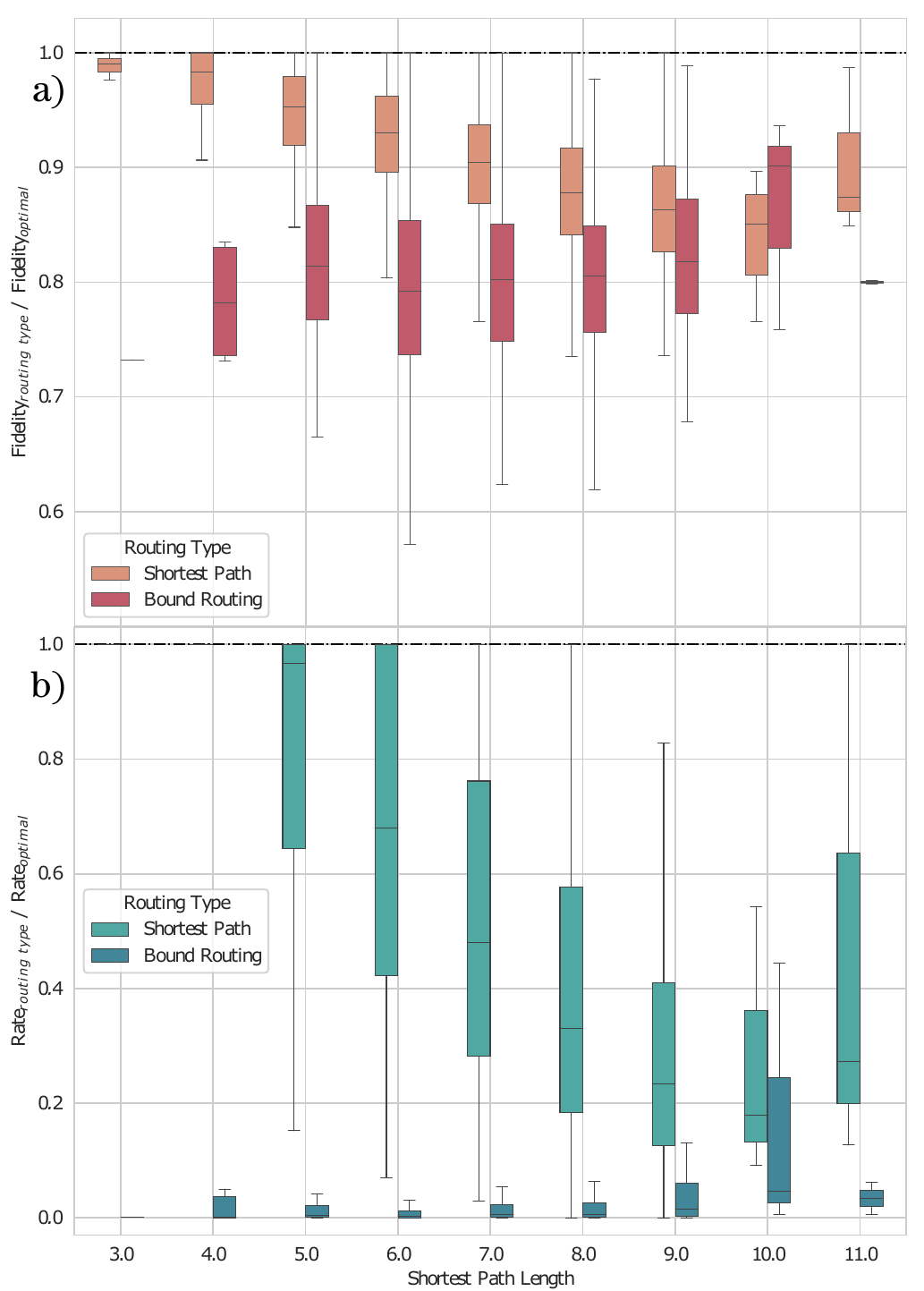}
    \caption{\color{black}Comparison between optimal routing for a 3-qubit GHZ state and shortest-path and bounds-based routing for the same state over Erdös-Rényi networks networks with 1000 nodes. The parameters utilized are: $\gamma_{min} = 0.9$, $p_\text{min} = 0.5$, $t_\text{min}=1$, $t_\text{max}=100$,  $\sigma_\text{min} = 10^4$, $\sigma_\text{max} =10^5$, $f_\text{trunc}^\text{GHZ}=0.5$, $f_\text{trunc}^\text{path} = 0.72$. Note that in this type of plot, the filled boxes contain the data between the Q1 and Q3 percentile, with the median being represented by a line, and the bar contains the minimum and the maximum of the distribution of points.\color{black}}
    \label{fig:internet}
\end{figure}
The second method is a \textit{bounds-based} method, following the work of \cite{Pirandola2019,Pirandola2019a,Pirandola2020,Bauml2020}, where we perform the routing based on the maximum bound on what a protocol could achieve on distributing GHZ states. To each edge, one attributes a capacity that translates the loss in the channel, and the probabilistic nature of entanglement generation, namely $C_{i:j} = p_{i:j} (1-H_2(3\gamma_{i:j}/4))$. Then, the capacity of distribution is bounded by the minimum capacity along the links that make a given path. Since the goal is to maximize the capacity of distribution, this is a max-min problem. To shorten some redundancy when two paths have the same minimum, we maximize the minimum capacity, and if two paths have the same capacity, we choose the one with the least amount of links. This guarantees a fair comparison with the other metrics.
In Fig. \ref{fig:internet} we plot the comparison between our optimal algorithm and routing under the above mentioned schemes. To do this, we perform the routing (optimization over the network) under the newly described metrics, and then calculate, using our derived metrics, the fidelity and rate for the solutions found. Note that the results never show a solution better than our optimal algorithm, and tend to be worse, as it was expected from the assumptions taken using those models of networks. Namely, for shortest-path routing, it is assumed homogeneity of the network and pure-states and, for bounds-based routing, it is assumed an information theoretical bound where the protocols that run at the ultimate limit that can be achieved for repeater-assisted quantum communications, which is less perceptive to how many links are used, as it functions as a bottleneck measure. When dealing with protocols where the fidelity decreases as the number of consumed pairs increases, this immediately creates a strong dependence on the size of the paths. For that reason, bounds routing achieves a comparatively worse result than shortest-path. In these plots, we used Erdös-Rényi networks with 1000 nodes. In Appendix \ref{appendix:optimalitysimul} we present simulations made on models of classical internet-like networks, and provide the conditions and simulations that retrieve the results of routing under the model of Ref.\cite{Meignant2018}.

\color{black}

\section{Conclusions} 

With this work, we provided a methodology and an algorithm capable of optimizing multipartite entanglement distribution over noisy quantum networks. To achieve this, we calculated the necessary metrics: \textit{(i)} the fidelity of the final state, depending on the noise and decoherence of the quantum memories, and \textit{(ii)} the rate of distribution, which depends on the communication time required and the stochastic behavior of the processes involved. Our methodology can be used to optimize additional parameters, as long as the metrics verify the properties of monotonicity, isotonicity and label-isotonicity. When extending to higher numbers of qubits, under a star-scheme, our algorithm is still optimal if we allow the links to be used more than once. Furthermore, by separating our approach into two equivalent layers, described by distinct algebras for routing and for trees, we allow bipartite entanglement distribution to be different from multipartite schemes. This opens the possibility for different protocols or algorithms within each layer and a condition that guarantees they fit together: label-isotonicity.

An additional important detail, which is exemplified by the chosen metrics for bipartite entanglement distribution, is that, while a set of parameters might not be isotonic, a decomposition into another set of isotonic parameters might still be possible. If this decomposition results in a larger number of parameters, as it did in our case, by increasing the number of objectives of optimization, the algorithms runtime also increases. This is partly derived from the fact that the set of optimal solutions grows. Nonetheless, doing so is compatible with our methodology, still providing a guarantee of optimality.

This same methodology could then be applied to different models of networks, for example having some nodes capable of distributing multipartite entanglement \cite{Avis2022a},  or different distribution schemes \cite{Wallnofer2016}, introducing entanglement purification rounds in the bipartite distribution scheme \cite{Goodenough2020}, or even different quantum repeater protocols \cite{Satoh2016}, \color{black} providing a starting point to optimally generate multipartite entanglement \color{black} over noisy quantum networks.

Given the development of networked quantum technologies, both over a future quantum internet, and over QLANs, optimizing the distribution of multipartite entanglement will naturally impact the deployment of the communications, computation and sensing applications that use this important resource \cite{Sekatski2019,Shettell2021}.
Future directions encompass using this algorithm with other metrics adapted to their physical realization of quantum networks or using this description of the metrics to find and verify the optimality of new algorithms for the multi-objective Steiner tree problem. While optimality might not always be the practical goal, these tools can be used to find a measure of closeness to optimality and possibly new routing protocols that are close to optimal. The broadness of this methodology should make for a robust tool to deal with routing on quantum networks.

\begin{acknowledgments}
The authors thank D. Markham and E. Z. Cruzeiro for useful comments on the work. Furthermore, the authors thank the support from Funda\c{c}\~{a}o para a Ci\^{e}ncia e a Tecnologia (Portugal), namely through projects UIDB/04540/2020 and UIDB/50008/2020, as well as from projects QuantHEP and HQCC supported by the EU QuantERA ERA-NET Cofund in Quantum Technologies and by FCT (QuantERA/0001/2019 and QuantERA/004/2021, respectively), and from the EU Quantum Flagship projects QIA (820445) and QMiCS (820505). B. Coutinho also thanks the support from FCT through project CEECINST/00117/2018/CP1495/{CT0001}.
\end{acknowledgments}

\onecolumn\newpage
\appendix

\section{Bipartite Entanglement Distribution - Metrics Properties \label{appendix:bipartite}}

As explained on the main text, the four parameters $[ p_{m:n}, t_{m:n},\gamma_{m:n},\sigma_{m:n}]$ for bipartite entanglement distribution should be routed for independently. This way, we can guarantee that the contraction from these four parameters into only three parameters detailed in the main text provides the optimal solutions. This is only true since every of these metrics are both monotone and isotone, which can be verified in the following set of inequalities for the:
\begin{enumerate}
    \item Probability of success metric
\begin{equation}
    \begin{aligned}
        &p_{m:n} \geq p_{m:n} \oplus (p_{n:o},k_n) = p_{m:n} \cdot p_{n:o} \cdot k_n  \\
        &p_{m:n}^{(1)} \geq p_{m:n}^{(2)} \ \Rightarrow \ p_{m:n}^{(1)} \oplus (p_{n:o},k_n) \geq p_{m:n}^{(2)} \oplus (p_{n:o},k_n)
    \end{aligned}
\end{equation}
$\forall p_{m:n}, p_{n:o}, k_n \in [0,1]$.
    \item Communications time metric
\begin{equation}
    \begin{aligned}
        &t_{m:n} \leq t_{m:n} \oplus t_{n:o} = t_{m:n} + 2 t_{n:o} \\
        &t_{m:n}^{(1)} \leq t_{m:n}^{(2)} \ \Rightarrow \ t_{m:n}^{(1)} \oplus t_{n:o} \leq t_{m:n}^{(2)} \oplus t_{n:o}
    \end{aligned}
\end{equation}
$\forall t_{m:n}, t_{n:o} \in \mathbb{R}_0^+$.
    \item Link fidelity metric
    \begin{equation}
    \begin{aligned}
        &\gamma_{m:n} \geq \gamma_{m:n} \oplus \gamma_{n:o} = \gamma_{m:n} \cdot \gamma_{n:o} \\
        &\gamma_{m:n}^{(1)} \geq \gamma_{m:n}^{(2)} \ \Rightarrow \ \gamma_{m:n}^{(1)} \oplus \gamma_{n:o} \geq \gamma_{m:n}^{(2)} \oplus \gamma_{n:o}
    \end{aligned}
    \end{equation}
$\forall \gamma_{m:n}, \gamma_{n:o} \in [0,1]$.
    \item and Decoherence time metric
    \begin{equation}
    \begin{aligned}
        &1/\sigma_{m:n} \leq 1/(\sigma_{m:n} \oplus \sigma_{n:o}) = 1/\sigma_{m:n} + 1/\sigma_{n:o} \\
        &1/\sigma_{m:n}^{(1)} \leq 1/\sigma_{m:n}^{(2)} \ \Rightarrow \ 1/(\sigma_{m:n}^{(1)} \oplus \sigma_{n:o}) \leq 1/(\sigma_{m:n}^{(2)} \oplus \sigma_{n:o})
    \end{aligned}
\end{equation}
$\forall \sigma_{m:n}, \sigma_{n:o} \in \mathbb{R}^+$.
\end{enumerate}

In all the previous inequalities, the first corresponds to monotonicity and the second to isotonicity. When the metric is separable, as in all these cases, both properties are usually verified.

\section{Proposition \ref{th:treeshortest} Proof \label{appendix:proof}}

We will first present the proof for the case of one objective of routing, or only one algebra, using only its total order and then generalize for the case of an arbitrary number of objectives using the dominance relation. Moreover, we will do this for a star with an arbitrary number of terminals. 

\begin{prop}
For the shortest-star with $3$ terminals, the paths connecting the center node and the terminals must be the shortest-paths, if the underlying algebras for trees are label-isotone.
\label{th:treeshortest}
\end{prop}

\begin{proof}
Consider the more general case in which the center node is connected by $n$ paths (this does not happen if the center node is one of the terminals, but for that case consider the star composed of $n-1$ paths), indexed by a number between 1 and $n$: $path_1, path_2 , ... path_n$ $\in \Sigma$. Each path is connected to one of $n$ terminals. Fix all paths but $path_1$.  Let $t \in \Xi$ be correspondent to the tree formed by $path_2 \cup path_3 \cup ... \cup path_n$. Now consider there $\exists \ \overline{path_1}: \overline{path_1} \preceq path_1$, due to label-isotonicity of the algebra for trees, then if $\overline{path_1} \preceq path_1 \Rightarrow f( t \oplus \overline{path_1} ) \preceq f( t \oplus path_1 )$ and the shortest tree would be $\overline{path_1} \cup path_2 \cup path_3 \cup ... \cup path_n $. Doing this for every other path, we get that the shortest-star is the one with every branch being the shortest-path between the center node and the terminals.
\end{proof}

\theoremstyle{remark}
\begin{remark}
The same applies for the multi-objective shortest-star problem, with the paths being the set of Pareto-optimal paths and requiring that every algebra for trees is label-isotone.
\end{remark}

\begin{proof}
Consider that $path_1 \notin X_1$ where $X_1$ is the set of Pareto-optimal paths between the node 1 and the center node. Then, $\exists \ \overline{path_1} \in X_1$ such that $\overline{path_1} \text{ D } path_1$ and from here the star containing $\overline{path_1}$ is better than the one containing $path_1$. This comes from the way the dominance relation is defined. The rest of the proof is identical to the previous one. 
\end{proof}

Moreover, further refinements in the search process can be obtained, taking advantage of properties of the algebras. Namely, if the algebra for trees is monotone, then a necessary, but not sufficient, condition for existence of a shortest-tree connecting the set of terminal can be obtained: if there is no possible path connecting some pair of terminals, then there is no solution of the problem. From the structure of the algorithm, this condition can be implemented while finding the shortest-paths from each terminal node. 

One important consideration is that the algebras for trees and the algebras for routing (parameters for trees and parameters for paths) do not need to be equal in number nor in form. As long as each algebra for routing is individually monotone and isotone, then the set of paths are optimal under an appropriate algorithm, which we will exemplify in \emph{Appendix \ref{appendix:mosp}}. And if each algebra for trees is label-isotone with respect to each individual algebra for routing it depends on, then this optimality is guaranteed.

Notice that in this proof we do not concern with different paths having intersections, as for the case with only three terminals, $T=3$, this is never the case. For $T>3$ our approach is only optimal if we allow intersections, which is equivalent to letting the same link be used more than once. If that is not the case, we can only extract at least part of the solutions and flag if there might exist more or not. If after running the algorithm and finding all possible solutions, with none of them having intersections, then all the optimal solutions were found. However, if some have intersections, then we can discard them (this is what we mean when we say \textit{a posteriori}) but we cannot guarantee all optimal solutions were found, since we ignored some solutions with non-optimal paths that did not have intersections. This can be translated in the following way: consider there exists a star $s = s_1 \oplus p_1$ such that $s_1 \cap p_1 \neq \emptyset$, this is, they share a common link, and that $s \in SA \equiv $ algorithm solutions. Now, consider there $\exists \ \tilde{p}_1 : p_1 \text{ D } \tilde{p}_1$ and $\tilde{s} = s_1 \oplus \tilde{p}_1$ and $ s_1 \cap \tilde{p}_1 = \emptyset$, there is a chance of $\tilde{s}$ being optimal since it has no intersections, but never ends up in the algorithm solutions because it contains a dominated path $\tilde{p}_1$. If however, every solution in $SA$ has no intersections and the algorithm is optimal, it would mean that any star with a non-optimal path, intersecting or not, would be dominated by at least one of the solutions and therefore would not be optimal.

\section{MOSP Algorithm \label{appendix:mosp}}

The multi-objective shortest-path (MOSP) algorithm \cite{Martins1984} is a powerful tool to find the optimal path in the Pareto sense, which is then the input for our algorithm of distributing multipartite entanglement. We implemented it in the following way, finding every optimal path from the source to every other node of the network:

\begin{algorithm}[H]
\caption{Multi-Objective Shortest-Path (MOSP)} \label{multiobjectiverouting} 
    \begin{algorithmic}[1] % The number tells where the line numbering should start
        \Procedure{Shortest-Path}{$source$} \Comment{Finds the shortest path to every node from the source}
            \State $Nodes :=$ Set of nodes of the network, each with underlying list of paths $Paths_u$ initialized as empty;
            \State $A :=$ Set of visited nodes of the network initialized as empty;
            \State $B :=$ Set of nodes to visit ordered as a priority queue data structure, with priority defined by the dominance relation;
            \State Initialize $source \gets \{ e_{\Sigma_i} \}$; 
            \State Add $source$ to $B$;
            \While{$B \neq $ empty}
            	\State $node \gets$ Top(B) 
        		\State Remove $node$ from B and add to A;
        		\For{$v \in$ neighbors($node$)}
        		    \State $Paths^{add} \gets$ possible paths from $\{Paths^{(i)}_{node} \oplus Edge(node,v)\}$;
        			\State Update $Paths_v$ considering $Paths^{add}$
        			\If{$Paths_v$ changes}
        			    \State Add/Update $v$ to/in B
        			    \If{$v \in $ A}
        			        \State Remove $v$ from A
        			    \EndIf
        			\EndIf
               	\EndFor
            \EndWhile
        \EndProcedure
    \end{algorithmic}
\end{algorithm}

\hspace{-0.4cm}where $\{ e_{\Sigma_i} \}$ are the neutral elements of $\Sigma_i$ w.r.t $\oplus_i$. The main reason to separate the fidelity metric 
\begin{equation}
    \frac{4F_{m:n}-1}{3} = \gamma_{m:n} e^{-t_{m:n}/\sigma_{m:n}}
\end{equation}
into three different metrics is that, while these parameters are individually monotonic and isotonic, together they are not. Then using this type of algorithm would fail at converging to the optimal solution that maximizes $F_{m:n}$. However, by separating them and treating them independently using the dominance relation, one can clearly see that, if one path dominates the other, then:
\begin{equation*}
        \left\{\begin{array}{c}
        \gamma_{m:n}^{(1)} \geq \gamma_{m:n}^{(2)}\\
        t_{m:n}^{(1)} \leq t_{m:n}^{(2)}\\
        \sigma_{m:n}^{(1)} \geq \sigma_{m:n}^{(2)}
        \end{array}\right\} \Rightarrow \  F_{m:n}^{(1)} \geq F_{m:n}^{(2)}
\end{equation*}
If not all partial relations on the left hand side are true, then the paths do not dominate each other and we cannot guarantee that, at every extension, the relation of the final fidelities would maintain. This requires keeping track of all the non-dominated paths -- which constitute the set of Pareto optimal paths.

\section{Tree-Scheme \label{appendix:treescheme}}
\begin{figure}[!ht]
\centering
\includegraphics[width=0.6\linewidth]{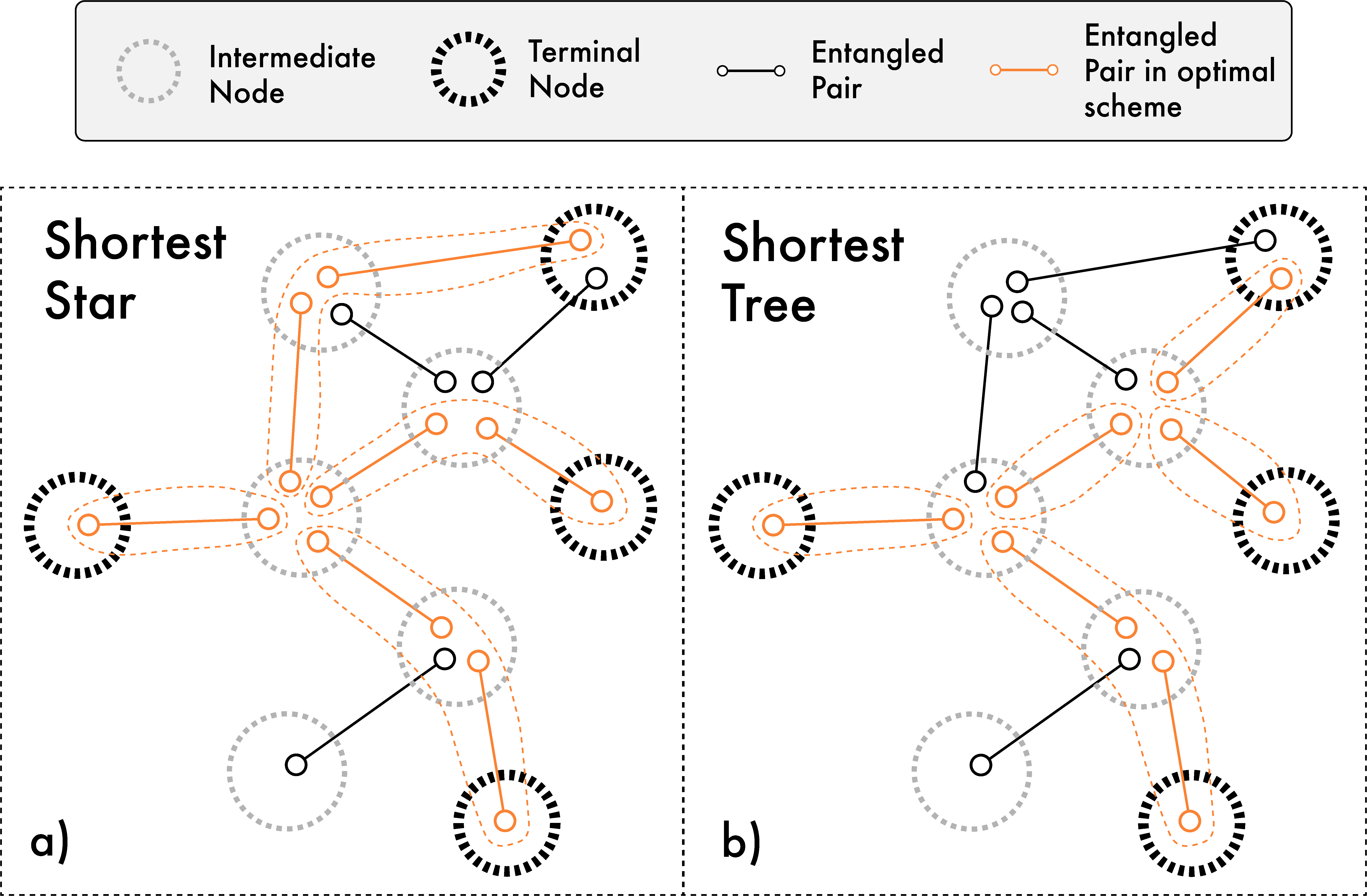}
\caption{A example of what a shortest tree and a shortest star are (highlighted in orange). Note that groups, or paths, surrounded by a traced orange line are equivalent to an entangled pair after bipartite entanglement distribution.}
\label{fig:shortest_treestar}
\vspace{-0.5cm}
\end{figure}

The tree scheme \cite{Meignant2018} consists of a generalisation of the star scheme, when we allow the topology of distribution to be any tree, not just a star-graph. Starting from a tree-graph connecting all terminal nodes  (see \emph{Figure \ref{fig:shortest_treestar}b}), choosing a leaf node and iteratively applying the star expansion protocol \cite{Meignant2018} for every node will result in GHZ state distributed across the terminal nodes. Under this distribution scheme, finding the optimal way to distribute entanglement across the network is equivalent to finding the shortest-tree connecting the terminal nodes, under a given set of parameters. This is translatable to solving the multi-objective Steiner tree algorithm with the corresponding metrics. The regular Steiner tree problem is known to be $NP$-Hard \cite{Wang2008,Robins2005} and adding the multi-objective setup quickly escalates the problem complexity. On the other hand, finding the shortest-star connecting a set of terminal nodes is a more tractable problem, and a reason why for 3-qubits, this problem is feasible.

To find the metrics expression for the $n$-qubits GHZ state distributed in a tree-scheme, we considered one center of coordination in order to minimize the communications time. This center of coordination, in the case of the star scheme was trivially the center node. Now, since the coordinating node is not necessarily one branch away from every terminal node, we have to perform a minimisation over the possible locations for this node:

\begin{align}
	t_{tree} = 2\cdot \frac{\underset{s \in \mathcal{S}}{\min}  \Big\{ \underset{\tau \in \mathcal{T}}{\max} \{ t_{s:\tau} \} \Big\} }{\underset{{m:n \in \text{branches}}}{\prod}p_{m:n}} \ , \ \xi_{tree} = \nicefrac{1}{t_{tree}}
	\label{eq:ratetree}
\end{align}
where $\mathcal{S}$ is the set of the Steiner nodes, $i.e.$, the set of nodes that belong to the tree, but are not terminal nodes. In our case, they are always intersections of branches. Moreover, notice that the numerator of $t_{tree}$ is the function that, given a tree, outputs the radius of the tree, under the communications time metric. 

Moving to the fidelity and density matrix of the distributed state: we must start from one terminal node, $\tau_0$, making the result dependent from which terminal we apply the procedure. This does not invalidate the scheme described when calculating the rate, since this only affects which operations are made in each node. Starting from an initial node, after performing the star-expansion protocol across all of the intermediary nodes of the tree, the final form of the state is equivalent, up to qubit swap operations which leave the GHZ part invariant and the fidelity unaffected, to applying a depolarising channel to each terminal node except the initial:
\begin{equation}
	 \mathfrak{D}_{\mathcal{T}_0 } (\cdot ) = \underset{\tau \in \mathcal{T} \setminus \tau_0}{\bigcirc} \mathcal{D}_{\tau}^{F_\tau} (\cdot) \triangleq \mathcal{D}_{\tau_1}^{F_{\tau_1}} \circ \mathcal{D}_{\tau_2}^{F_{\tau_2}} \circ ... 
\end{equation}
and applying, for each Steiner node, a depolarising channel to the initial node:
\begin{equation}
\begin{aligned}
	\mathfrak{D}_{\mathcal{S}_0} (\cdot )  = \underset{s \in \mathcal{S} }{\bigcirc} \mathcal{D}_{\tau_0}^{F_s} (\cdot) \triangleq \mathcal{D}_{\tau_0}^{F_{s_0}} \circ \mathcal{D}_{\tau_0}^{F_{s_1}} \circ ... \\
	\label{eq:steinernodes}
\end{aligned}
\end{equation}
Making the final result:
\begin{equation}
\begin{aligned}
    \mathfrak{D}_{\mathcal{T}_0 } \circ \mathfrak{D}_{\mathcal{S}_0}   \Big( \ket{GHZ}\bra{GHZ} \Big) .
\end{aligned}
\label{eq:depall}
\end{equation}

In all the above, $\circ$ stands for composition and ${\bigcirc}_{a \in \mathcal{A} }$ represents the enumerated composition for every $a$ in the set $\mathcal{A}$. The fidelities of each depolarising channel correspond to the fidelities of each branch connecting either the terminals or Steiner nodes, respectively. The final fidelity of this state can be calculated, with the second description of the depolarising channel, given in the main text, revealing again to be particularly useful. 
Separating the calculations by the different entries of the density matrix of an GHZ state, let us start on the diagonal terms, dismissing the non-GHZ entries:

\begin{equation}
\begin{aligned}
    \mathcal{D}_{\tau_0}^{F_{s_0}} \circ \mathfrak{D}_{\mathcal{T}_0 } ( \ket{\mathbf{0}}\bra{\mathbf{0}} ) = \prod_{\tau \in \mathcal{T} \setminus \tau_0 \cup s_0} \frac{1+2F_{\tau}}{3} \ket{\mathbf{0}}\bra{\mathbf{0}} + ...\\
    ... + \prod_{\tau \in \mathcal{T} \setminus \tau_0 \cup s_0} \frac{2(1-F_{\tau})}{3} \ket{\mathbf{1}}\bra{\mathbf{1}}
\end{aligned},
\label{eq:entry00}
\end{equation}

\begin{equation}
\begin{aligned}
    \mathcal{D}_{\tau_0}^{F_{s_0}} \circ \mathfrak{D}_{\mathcal{T}_0 } ( \ket{\mathbf{1}}\bra{\mathbf{1}} ) = \prod_{\tau \in \mathcal{T} \setminus \tau_0 \cup s_0} \frac{1+2F_{\tau}}{3} \ket{\mathbf{1}}\bra{\mathbf{1}} + ...\\
    ... + \prod_{\tau \in \mathcal{T} \setminus \tau_0 \cup s_0} \frac{2(1-F_{\tau})}{3} \ket{\mathbf{0}}\bra{\mathbf{0}}
\end{aligned},
\label{eq:entry11}
\end{equation}
with $\ket{\mathbf{0}}\bra{\mathbf{0}}$ and $\ket{\mathbf{1}}\bra{\mathbf{1}}$ denoting the density matrix entries $\ket{00...0}\bra{00..0}$ and $\ket{11...1}\bra{11...1}$, respectively. The off-diagonal terms are actually eigenvectors, with eigenvalue $(4F_i-1)/3$, of the depolarising channel, making the calculations very straightforward:

\begin{equation}
\begin{aligned}
    \mathcal{D}_{\tau_0}^{F_{s_0}} \circ \mathfrak{D}_{\mathcal{T}_0 } ( \ket{\mathbf{0}}\bra{\mathbf{1}} ) = \prod_{\tau \in \mathcal{T} \setminus \tau_0 \cup s_0} \frac{4F_{\tau}-1}{3} \ket{\mathbf{0}}\bra{\mathbf{1}} ,
\end{aligned}
\label{eq:entry01}
\end{equation}
\begin{equation}
\begin{aligned}
    \mathcal{D}_{\tau_0}^{F_{s_0}} \circ \mathfrak{D}_{\mathcal{T}_0 } ( \ket{\mathbf{1}}\bra{\mathbf{0}} ) = \prod_{\tau \in \mathcal{T} \setminus \tau_0 \cup s_0} \frac{4F_{\tau}-1}{3} \ket{\mathbf{1}}\bra{\mathbf{0}} .
\end{aligned}
\label{eq:entry10}
\end{equation}

The next step in the calculation is adding the Steiner nodes' depolarising channels. For this, let us first introduce two complementary functions, $E(\mathcal{S})$ and $O(\mathcal{S})$, that translate even and odd applications of $2(1-F_i)/3 \cdot \Lambda_{\tau_0} (\hat{Y}_{\tau_0} \rho \hat{Y}_{\tau_0}^\dagger)$. Throughout the rest of this section, let $\mathcal{S}$ stand for the vector of fidelities of branches connecting to the Steiner nodes $\{ F_{s_1}, F_{s_2} ... \}$ and adding branches is equivalent to $\{ F_{s_1}, F_{s_2},...,F_{s_n}\} \oplus \{ F_s \} = \{ F_{s_1}, F_{s_2},...,F_{s_n}, F_s\} $. Also, for simplicity define, $\mathcal{F}_i := (1+2F_i)/3$ and $\overline{\mathcal{F}}_i := 1-\mathcal{F}_i = 2(1-F_i)/3$. 

A simple example of how to calculate this can be made, using \emph{Eq. \ref{eq:steinernodes}} with only two Steiner nodes:
\begin{equation}
\begin{aligned}
	\mathfrak{D}_{\mathcal{S}_0} ( \rho )  &=  \mathcal{D}_{\tau_0}^{F_{s_1}} \circ \mathcal{D}_{\tau_0}^{F_{s_2}} (\rho) \\
	&= \mathcal{D}_{\tau_0}^{F_{s_1}} \left( \mathcal{F}_{s_2} \  \rho + \overline{\mathcal{F}}_{s_2}  \Lambda_0 (\hat{Y}_0 \rho \hat{Y}_0^\dagger)\right) \\
	&= \left( \mathcal{F}_{s_1} \cdot \mathcal{F}_{s_2}  + \overline{\mathcal{F}}_{s_1} \cdot \overline{\mathcal{F}}_{s_2} \right) \rho \ + \\
	& \quad + \left( \mathcal{F}_{s_1}\cdot \overline{\mathcal{F}}_{s_2} +  \overline{\mathcal{F}}_{s_1} \cdot \mathcal{F}_{s_2} \right) \Lambda_0 (\hat{Y}_0 \rho \hat{Y}_0^\dagger) \\
	&\triangleq E\left(\{ F_{s_1},  F_{s_2}\}\right) \rho + O(\{ F_{s_1},  F_{s_2}\})  \Lambda_0 (\hat{Y}_0 \rho \hat{Y}_0^\dagger)
\end{aligned}.
\end{equation}
We can completely define the functions recursively by:
\begin{equation}
    \begin{cases}
        E(\{ F_{s_0} \} )= \mathcal{F}_{s_0}    \\
        O(\mathcal{S}) = 1-E(\mathcal{S}) \\
       E(\mathcal{S}\oplus s) = E(\mathcal{S}) \cdot \mathcal{F}_s + O(\mathcal{S}) \cdot \overline{\mathcal{F}}_s \\
       O(\mathcal{S}\oplus s) = O(\mathcal{S}) \cdot \mathcal{F}_s + E(\mathcal{S}) \cdot \overline{\mathcal{F}}_s \\
    \end{cases},
    \label{eq:propertiesEO}
\end{equation}
and find some of its properties, namely:
\begin{equation}
    \begin{cases}
       E(\mathcal{S}) \geq O(\mathcal{S}) \qquad, \text{if } \forall i: F_{s_i} > 1/2\\
       E(\mathcal{S}) - E(\mathcal{S}\oplus s) = [E(\mathcal{S}) - O(\mathcal{S})]\cdot \overline{\mathcal{F}}_s \\
       O(\mathcal{S}\oplus s) - O(\mathcal{S}) = [E(\mathcal{S}) - O(\mathcal{S})]\cdot \overline{\mathcal{F}}_s
    \end{cases},
    \label{eq:propertiesE1}
\end{equation}
where $s \equiv \{F_s \}$. Adding to \emph{Eqs. \ref{eq:entry00}, \ref{eq:entry11}, \ref{eq:entry01}, \ref{eq:entry10}}, the missing depolarising channels of \emph{Eq. \ref{eq:depall}}, they become:
\begin{equation}
\begin{aligned}
    \mathfrak{D}_{\mathcal{T}_0 } \circ \mathfrak{D}_{\mathcal{S}_0}   ( \ket{\mathbf{0}}\bra{\mathbf{0}} ) = E(\mathcal{S}) \prod_{\tau \in \mathcal{T} \setminus \tau_0} \frac{1+2F_{\tau}}{3} \ket{\mathbf{0}}\bra{\mathbf{0}} + ...\\
    ... + O(\mathcal{S}) \prod_{\tau \in \mathcal{T} \setminus \tau_0 } \frac{2(1-F_{\tau})}{3} \ket{\mathbf{1}}\bra{\mathbf{1}}
\end{aligned},
\end{equation}
\begin{equation}
\begin{aligned}
    \mathfrak{D}_{\mathcal{T}_0 } \circ \mathfrak{D}_{\mathcal{S}_0}   ( \ket{\mathbf{1}}\bra{\mathbf{1}} ) = E(\mathcal{S}) \prod_{\tau \in \mathcal{T} \setminus \tau_0 } \frac{1+2F_{\tau}}{3} \ket{\mathbf{1}}\bra{\mathbf{1}} + ...\\
    ... + O(\mathcal{S}) \prod_{\tau \in \mathcal{T} \setminus \tau_0 } \frac{2(1-F_{\tau})}{3} \ket{\mathbf{0}}\bra{\mathbf{0}}
\end{aligned},
\end{equation}
\begin{equation}
\begin{aligned}
    \mathfrak{D}_{\mathcal{T}_0 } \circ \mathfrak{D}_{\mathcal{S}_0}  ( \ket{\mathbf{0}}\bra{\mathbf{1}} ) = \prod_{\tau \in \mathcal{T} \setminus \tau_0 \cup \mathcal{S}} \frac{4F_{\tau}-1}{3} \ket{\mathbf{0}}\bra{\mathbf{1}} 
\end{aligned},
\end{equation}
\begin{equation}
\begin{aligned}
    \mathfrak{D}_{\mathcal{T}_0 } \circ \mathfrak{D}_{\mathcal{S}_0}   ( \ket{\mathbf{1}}\bra{\mathbf{0}} ) = \prod_{\tau \in \mathcal{T} \setminus \tau_0 \cup \mathcal{S}} \frac{4F_{\tau}-1}{3} \ket{\mathbf{1}}\bra{\mathbf{0}} 
\end{aligned}.
\end{equation}

Hence, projecting the final state over a GHZ state, the final result for the fidelity becomes:

\begin{equation}
\begin{aligned}
	f = \frac{1}{2} \Bigg[ E(\mathcal{S})  \prod_{\tau \in \mathcal{T} \setminus \tau_0} \frac{1+2F_{\tau}}{3} + O(\mathcal{S}) \prod_{\tau \in \mathcal{T} \setminus \tau_0} \frac{2(1-F_{\tau})}{3}  + \prod_{s \in \mathcal{S}} \frac{4F_{s} -1}{3} \cdot \prod_{\tau \in \mathcal{T} \setminus \tau_0} \frac{4F_{\tau} -1}{3} \Bigg],
	  \label{eq:fidelitytree}
\end{aligned}
\end{equation}

Moreover, since every star is a tree, these metrics also apply for the \textit{star-scheme} under the correct assumptions. It is easy to verify that in a star scheme, there is only one Steiner node (the center node) which translates in the functions $E(\mathcal{S})$ and $O(\mathcal{S})$ taking the values $(1+2F_{\tau_0})/3$ and $2(1-F_{\tau_0})/3$ respectively.

\section{W-States Distribution \label{appendix:wstate}}

W States are a well known class of multipartite entangled states. As pointed in \cite{Dur2000}, for 3-qubits, W states, together with GHZ states, compose the only two inequivalent classes of multipartite entangled states. For this reason we also include them here. Since the tree scheme introduced in Appendix \ref{appendix:treescheme} only works for GHZ states, the only scheme we analyse here for the distribution of W states is the usual star scheme introduced in the main text, where after the distribution of bipartite entanglement between all the terminal nodes and the center node, one would have to perform a projection of a W state in the qubits located in the center node. This would distribute the W state across the terminal nodes' qubits.

For this reason, the only metric that changes in the star scheme for the distribution of the W state is the fidelity metric which explicitly depends on the state, and, more exactly, on how the depolarising channel acts on the quantum state. To calculate the exact final fidelity metric let us first introduce the following notation we use henceforth to write the W state with $n$ qubits:
\begin{equation}
\begin{aligned}
        \ket{W} &= \frac{1}{n} \sum_{i=1}^n \ket{\underline{i}} 
\end{aligned},
\end{equation}
where $\ket{\underline{i}} := \left( \bigotimes_{j \neq i} \ket{0}_j \right) \otimes \ket{1}_i =  \ket{00...01_i0...0}$. Using this notation, let us describe the action of the depolarising channel onto each of the W states matrix entries $\ket{\underline{i}}\bra{\underline{j}}$, as we did for the GHZ state. Using Eq. \ref{eq:DepolarisingChannel}, we separate this channel into its two distinct Kraus operators: the first operator acts trivially on any of the entries of the matrix, leaving us only with the following operator to analyse:
\begin{equation}
    \Lambda_{\tau} ( \hat{Y}_\tau \ket{\underline{i}}\bra{\underline{j}} \hat{Y}_\tau^\dagger) =
    \begin{cases}
       \ket{\mathbf{0}}\bra{\mathbf{0}} &, \tau = i = j \\
       -\ket{\underline{i}}\bra{\underline{j}} &, i\neq j , i = \tau \vee j = \tau\\
       \ket{\underline{i},\underline{\tau}}\bra{\underline{j},\underline{\tau}} &,  i \neq \tau \wedge j \neq \tau
    \end{cases},
\end{equation}
apart from the constant $2(1-F_\tau)/3$ and where we denote the states $\ket{\mathbf{0}} = \ket{00...0}$ and $\ket{\underline{i},\underline{j}} = \ket{00...1_i...1_j...0}$. Notice that both the state $\ket{\mathbf{0}}\bra{\mathbf{0}}$ and the state $\ket{\underline{i},\underline{\tau}}\bra{\underline{j},\underline{\tau}}$ do never account for the fidelity. Moreover, after applying a depolarising channel on another qubit $\eta$, the entry $\ket{\mathbf{0}}\bra{\mathbf{0}}$ can contribute for the fidelity, whereas the same can never be said for $\ket{\underline{i},\underline{\tau}}\bra{\underline{j},\underline{\tau}}$, unless $\eta = \tau$, which is never the case.

Using these results, let us separate the terms that come from the diagonal entries and the terms that come from the off-diagonal entries. Consider that:
\begin{equation}
    \mathfrak{D}_{\mathcal{T} } (\cdot ) = \underset{\tau \in \mathcal{T} }{\bigcirc} \mathcal{D}_{\tau}^{F_\tau} (\cdot) \triangleq \mathcal{D}_{\tau_1}^{F_{\tau_1}} \circ \mathcal{D}_{\tau_2}^{F_{\tau_2}} \circ ... 
\end{equation}
Then, after applying all the $n$ depolarisation channels to the W state, we can derive the following expressions that add up to the final fidelity of the teleported state:

\begin{align}
    \mathfrak{D}_{\mathcal{T} } (\ket{\underline{\eta}}\bra{\underline{\eta}}) &= \left[ \prod_{\tau \in \mathcal{T}} \frac{1+2F_\tau}{3} + \sum_{\theta \in \mathcal{T}} \frac{2(1-F_\eta)}{3} \cdot  \frac{2(1-F_\theta)}{3} \cdot \prod\limits_{\substack{\tau \in \mathcal{T} \\ \tau \neq \theta,\eta}} \frac{1+2F_\tau}{3}   \right] \ket{\underline{\eta}}\bra{\underline{\eta}} \\
    \mathfrak{D}_{\mathcal{T} } (\ket{\underline{\eta}}\bra{\underline{\mu}}) &= \left[ \frac{4F_\eta -1}{3} \cdot \frac{4F_\mu -1}{3} \cdot    \prod\limits_{\substack{\tau \in \mathcal{T} \\ \tau \neq \mu,\eta}} \frac{1+2F_\tau}{3}   \right] \ket{\underline{\eta}}\bra{\underline{\mu}} + ...
\end{align}
\begin{equation}
\begin{aligned}
    f &= \frac{1}{n} \prod_{\tau \in \mathcal{T} } \frac{1+2F_\tau}{3} + \frac{1}{n^2} \sum\limits_{\substack{\eta, \theta \in \mathcal{T} \\ \theta \neq \eta}} \left[ \frac{2(1-F_\eta)}{3}   \frac{2(1-F_\theta)}{3} + \frac{4F_\eta-1}{3}   \frac{4F_\theta-1}{3} \right] \cdot \prod\limits_{\substack{\tau \in \mathcal{T} \\ \tau \neq \theta,\eta}} \frac{1+2F_\tau}{3}  \\
    &= \prod_{\tau \in \mathcal{T} } \frac{1+2F_\tau}{3} + \frac{2(n-1)}{n^2} \sum\limits_{\substack{\eta \in \mathcal{T}}} \left[ \frac{2(1-F_\eta)}{3}  \prod\limits_{\substack{\tau \in \mathcal{T} \\ \tau \neq \eta}} \frac{1+2F_\tau}{3} \right]  + \frac{2}{n^2} \sum\limits_{\substack{\eta, \theta \in \mathcal{T} \\ \theta \neq \eta}} \left[ \frac{2(1-F_\eta)}{3}   \frac{2(1-F_\theta)}{3}  \prod\limits_{\substack{\tau \in \mathcal{T} \\ \tau \neq \theta,\eta}} \frac{1+2F_\tau}{3} \right] \\
    &\equiv \prod_{\tau \in \mathcal{T} } \mathcal{F}_\tau + \frac{2(n-1)}{n^2} \sum\limits_{\substack{\eta \in \mathcal{T}}} \left[ \overline{\mathcal{F}}_\eta \prod\limits_{\substack{\tau \in \mathcal{T} \\ \tau \neq \eta}} \mathcal{F}_\tau \right]  + \frac{2}{n^2} \sum\limits_{\substack{\eta, \theta \in \mathcal{T} \\ \theta \neq \eta}} \left[ \overline{\mathcal{F}}_\eta   \overline{\mathcal{F}}_\theta  \prod\limits_{\substack{\tau \in \mathcal{T} \\ \tau \neq \theta,\eta}} \mathcal{F}_\tau \right] 
    \label{eq:wmetricfid}
\end{aligned},
\end{equation}
where we made use of the previously introduced notation $\mathcal{F}_i := (1+2F_i)/3$ and $\overline{\mathcal{F}}_i := 1-\mathcal{F}_i = 2(1-F_i)/3$.

\section{Monotonicity for GHZ States \label{appendix:monotonicity}}

In this section we will prove that every algebra for trees used is monotone, for both the \textit{star-scheme} and the \textit{tree-scheme} detailed previously. Starting with the fidelity for the more general GHZ state, under any of the possible schemes which are detailed in the main text, we have to prove that the addition of any path to the tree results always in a worse fidelity.

 By looking at the three different products in \emph{Eq. \ref{eq:fidelitytree}}, and separating them into a vector of length three, corresponding to the signature, one can then describe the correspondent algebra for trees:

$f_\text{GHZ}: \Big( [\nicefrac{1}{2};1) \cup \{0\} , \geq , (\nicefrac{1}{2};1) , (0;1)^3 , 0, \oplus_\text{GHZ}, h \Big)$ with $\oplus_\text{GHZ} : (0;1)^3 \times  (\nicefrac{1}{2};1)  \longrightarrow  (0;1)^3$ given by:

\begin{align}
	\begin{aligned}
		\big(\{ E(\mathcal{S})\cdot a , O(\mathcal{S})\cdot b , c \} &, \ F_{i} \big)  \mapsto 
	    \begin{cases}
		    \{ E(\mathcal{S})\cdot a \cdot \mathcal{F}_i ,O(\mathcal{S})\cdot  b \cdot \overline{\mathcal{F}}_i , c \cdot (\mathcal{F}_i-\overline{\mathcal{F}}_i) \} ,& i \cap \mathcal{T} \neq \emptyset\\
		    \{ E(\mathcal{S}\oplus \{ F_i \} )\cdot a ,O(\mathcal{S} \oplus \{ F_i \})\cdot  b , c \cdot (\mathcal{F}_i-\overline{\mathcal{F}}_i) \} ,& i \cap \mathcal{T} = \emptyset
		\end{cases}
	\end{aligned}
	\label{eq:extensiontree}
\end{align}

where $\{E(\mathcal{S}) \cdot a,O(\mathcal{S}) \cdot b,c\} \in (0;1)^3$ is a general signature of a tree. Looking at \emph{Eq. \ref{eq:fidelitytree}} and considering that fidelities below $1/2$ can be discarded, $h(\cdot)$ is given by:
\begin{equation}
	h(\{ a , b , c \}) = 
	\begin{cases}
		\frac{a+b+c}{2} \quad &\ ,\text{ if } \frac{a+b+c}{2} \geq \nicefrac{1}{2} \\
		0 \quad &\ ,\text{ if } \frac{a+b+c}{2} < \nicefrac{1}{2}
	\end{cases}
	\label{eq:hfunc}
\end{equation}

Now that we can fully characterize the algebra (or metric), let us prove the monotonicity property of this algebra. Considering one general tree, given by the following signature:

\begin{equation*}
    \{ E(\mathcal{S}) \cdot a, O(\mathcal{S})\cdot b, c \}.
\end{equation*}

After performing an extension with another path, as stated in \emph{Eq. \ref{eq:extensiontree}}, there are two possible options. Either the path connects to a terminal node $\tau$ or to a Steiner node $s$. In the case it connects to a terminal node, the metric is trivially monotonic since every element in the signature would be multiplied by a value smaller than one:

\begin{equation*}
        \left\{\begin{array}{c}
        E(\mathcal{S}) \cdot a\\
         O(\mathcal{S})\cdot b\\
        c
        \end{array}\right\}
    \longmapsto
    \left\{\begin{array}{c}
        E(\mathcal{S}) \cdot a \cdot \mathcal{F}_\tau\\
         O(\mathcal{S})\cdot b \cdot \overline{\mathcal{F}}_\tau\\
        c \cdot (\mathcal{F}_\tau-\overline{\mathcal{F}}_\tau)
        \end{array}\right\}
\end{equation*}
In the case the path connects to a Steiner node then we need to use the know properties for $E(\cdot)$ and $O(\cdot)$ from \emph{Eq. \ref{eq:propertiesE1}} and prove that:

\begin{equation}
\begin{aligned}
    &
    \begin{cases}
       E(\mathcal{S}\oplus s) \cdot a + O(\mathcal{S}\oplus s)\cdot b \leq  E(\mathcal{S})\cdot a + O(\mathcal{S}) \cdot b \\
       c \cdot (\mathcal{F}_s - \overline{\mathcal{F}}_s) \leq c
    \end{cases}.
\end{aligned}
\end{equation}

The first expression is always true, considering the properties of $E(\cdot)$ and $O(\cdot)$ and that for any tree made up of $n$ possible paths (paths with a fidelity $F_i > 1/2$), $0 \leq b \leq (\nicefrac{1}{3})^n < (\nicefrac{2}{3})^n \leq a \leq 1$ and $F_s-\overline{F_s} \leq 1$. The second expression is also always true as $\mathcal{F}_s - \overline{\mathcal{F}}_s \leq 1$. This proves the monotonicity of this metric.

Moving to the rate metric, consider again the more general case for the GHZ state under any of the possible schemes. The first thing to notice is that while performing a minimization over all nodes for placing the coordination center, one can never actually reduce the maximum of the communications time while adding another path. This comes from the fact that in a tree, if after adding another path, this coordination center position changes, it must change to a place that still has a larger communications time, or else the previous position was not the one for the minimum to be attainable. The corresponding algebra for trees then becomes:

$\xi_\text{GHZ}: \Big( \mathbb{R}^+_0 , \geq , (0,1)\times \mathbb{R}^+ , (0,1)\times \mathbb{R}^+ , 0, \oplus_{\xi}, g \Big)$ with $\oplus_{\xi} : \Big((0,1)\times \mathbb{R}^+\Big) \times \Big((0,1)\times \mathbb{R}^+\Big)  \longrightarrow (0,1)\times \mathbb{R}^+$ given by

\begin{align}
	\begin{aligned}
		(\{ p_{tree} , t_{tree}\},\{ p_{m:n} , t_{m:n}\} )  \longmapsto  \{ p_{tree} \cdot p_{m:n} , \text{Radius} (t_{tree} \oplus t_{m:n}) \} ,
	\end{aligned}
\end{align}
where $\{p_{tree} , t_{tree}\} \in (0,1)\times \mathbb{R}^+$ is a general signature and $\text{Radius} (t_{tree})$ is a function that retrieves the radius of a tree under the metric $t$, which in this case is the communications time metric. Then, using \emph{Eq. \ref{eq:ratetree}}, the algebra weight function, $g(\cdot)$, is given by

\begin{equation}
	g(\{ p_{tree} , t_{tree} \}) = \frac{p_{tree}}{2 t_{tree}}.
	\label{eq:rateg}
\end{equation}

To prove the monotonicity, let us start from a general signature for the rate metric $\{p_{tree} , t_{tree}\} \in (0,1)\times \mathbb{R}^+$, adding any path with a probability $0\leq p_{m:n} \leq 1$ and a waiting time $t_{m:n} > 0$ would result in:

\begin{equation}
\begin{aligned}
    \xi = \frac{p_{tree}}{2t_{tree}}\mapsto \xi^{'} = &\frac{p_{tree}\cdot p_{m:n}}{2 \cdot \text{Radius} (t_{tree} \oplus t_{m:n}) }  \\ 
    \leq& \frac{ p_{tree}}{2 \cdot \text{Radius} (t_{tree} \oplus t_{m:n}) }  \\
    \leq & \frac{ p_{tree}}{2 \cdot t_{tree} } .
\end{aligned}
\end{equation}

In either possible case the rate always decreases. This proves that both the fidelity metric and the rate metric for a GHZ state, under any of the possible schemes, is monotonic.

\section{Label-Isotonicity for GHZ States \label{appendix:label}}

In this section we will focus on the properties of the algebras for trees in a GHZ state distribution under the \textit{star-scheme}, namely in determining wether they are label-isotone with respect to every single algebra for routing they depend on. Starting with the fidelity metric, the correspondent algebra is identical to the one described in the previous section, but in a \textit{star-scheme}, since there is only one Steiner node, every branch connects to a terminal. From the fidelity metric on the main text, by separating the three different identical parts in a vector, we can arrive at the algebra: 

$f_\text{GHZ}: \Big( [\nicefrac{1}{2};1) \cup \{0\} , \geq , (\nicefrac{1}{2};1) , (0;1)^3 , 0, \oplus_\text{GHZ}, h \Big)$ with $\oplus_\text{GHZ} : (0;1)^3 \times  (\nicefrac{1}{2};1)  \longrightarrow  (0;1)^3$ given by:
\begin{align}
	\begin{aligned}
		(\{ a , b , c \} , \ F_{m:n} )  \longmapsto  \{ a \cdot \frac{1+2F_{m:n}}{3} , b \cdot \frac{2(1-F_{m:n})}{3} , c \cdot \frac{4F_{m:n}-1}{3} \} ,
	\end{aligned}
\end{align}
and, using the same arguments as in \emph{Eq. \ref{eq:hfunc}}, $h(\cdot)$ is given by
\begin{equation}
	h(\{ a , b , c \}) = 
	\begin{cases}
		\frac{a+b+c}{2} \quad &\ ,\text{ if } \frac{a+b+c}{2} \geq \nicefrac{1}{2} \\
		0 \quad &\ ,\text{ if } \frac{a+b+c}{2} < \nicefrac{1}{2}
	\end{cases}.
\end{equation}

Now, to prove the label-isotonicity itself, consider a tree $t$ with corresponding fidelity signature $\{a,b,c\} \in (0,1)^3$. Moreover, consider two different paths with fidelity signatures correspondent to $\sigma_1$, $\sigma_2$ such that $\sigma_1 \preceq \sigma_2$, $i.e.$, $\sigma_1$ has a fidelity given by $F_{m:n}^{(1)}$ and $\sigma_2$ has a fidelity given by $F_{m:n}^{(2)} \leq F_{m:n}^{(1)}$. Then:
\begin{align}
	\begin{aligned}
		&F_{m:n}^{(1)} \geq F_{m:n}^{(2)} \Rightarrow \\
		&\Rightarrow a\cdot \frac{1+2F_{m:n}^{(1)}}{3} + b\cdot \frac{2(1-F_{m:n}^{(1)})}{3} + c\cdot \frac{4F_{m:n}^{(1)}-1}{3} \\
		&=  \frac{a+2b-c}{3} + \frac{2a-2b+4c}{3}\cdot F_{m:n}^{(1)} \\
		&\geq \frac{a+2b-c}{3} + \frac{2a-2b+4c}{3}\cdot F_{m:n}^{(2)}\ ,
	\end{aligned}
\end{align}
$\forall a,b,c \in (0,1), a>b \ ; \ F_{m:n}^{(1)}, F_{m:n}^{(2)} \in (\nicefrac{1}{2},1)$, which is always the case. This proves the label-isotonicity of the fidelity of a GHZ state. 

As for the rate, the corresponding algebra for trees, under a star scheme, is given by:

$\xi_\text{GHZ}: \Big( \mathbb{R}^+_0 , \geq , (0,1)\times \mathbb{R}^+ , (0,1)\times \mathbb{R}^+ , 0, \oplus_{\xi}, g \Big)$ with $\oplus_{\xi} : \Big((0,1)\times \mathbb{R}^+\Big) \times \Big((0,1)\times \mathbb{R}^+\Big)  \longrightarrow (0,1)\times \mathbb{R}^+$ given by:

\begin{align}
	\begin{aligned}
		(\{ p_{star} , t_{star}\}, \{ p_{m:n} , t_{m:n}\} )  \longmapsto 
		\{ p_{star} \cdot p_{m:n} , \max (t_{star} , t_{m:n}) \} ,
	\end{aligned}
\end{align}
where $\{p_{star} , t_{star}\} \in (0,1)\times \mathbb{R}^+$ are again a general signature. Then, using similar arguments as in \emph{Eq. \ref{eq:rateg}}, $g$ in this case is given by:

\begin{equation}
	g(\{ p_{star} , t_{star}\}) = \frac{p_{star} }{2 t_{star}}.
\end{equation}

As one can see, this algebra actually depends on two parameters from each path: the probability of success and the communications time. For this reason, we have to guarantee that the algebra is label-isotone with respect to each parameter in order for \emph{Proposition \ref{th:treeshortest}} to work. Starting with the probability of success, consider two paths with corresponding probabilities $p_{m:n}^{(1)}$ and $p_{m:n}^{(2)}$ such that:

\begin{align}
	\begin{aligned}
		p_{m:n}^{(1)} \geq p_{m:n}^{(2)} \Rightarrow  \frac{p_{star}\cdot p_{m:n}^{(1)}}{2 t_{star}} \geq  \frac{p_{star}\cdot p_{m:n}^{(2)}}{2t_{star}} \ , \quad \forall \ p_{star} \in (0,1),  \forall \ t_{star} \in \mathbb{R}^+. 
	\end{aligned}
\end{align}
Moving to the communications time, consider two paths with corresponding communication times $t_{m:n}^{(1)}$ and $t_{m:n}^{(2)}$ such that:
\begin{align}
	\begin{aligned}
		t_{m:n}^{(1)} \leq t_{m:n}^{(2)} \Rightarrow  \frac{p_{star}}{2\max(t_{star}, t_{m:n}^{(1)})} \geq  \frac{p_{star}}{2 \max(t_{star} , t_{m:n}^{(2)})} \ , \quad \forall \ p_{star} \in (0,1),  \forall \ t_{star} \in \mathbb{R}^+.
	\end{aligned}
\end{align}
This guarantees that both metrics are label-isotonic with respect to the corresponding algebras for routing, allowing the dominance relation used for paths to be cohesive with the dominance relation used for trees and ensuring the optimality of our algorithm.

\section{Monotonicity and Label-Isotonicity for W States \label{appendix:Wproperties}}

In this section we prove both the monotonicity and label-isotonicity for the fidelity metric of a W state. We prove them independently, as in this case, by making the fidelity of one of the $n$ paths go to 1, we do not retrieve the expression for the fidelity of $n-1$-qubits W state.
Let us first try to describe the fidelity metric into an algebra for trees:

$f_\text{W}: \Big( [\nicefrac{1}{2};1) \cup \{0\} , \geq , (\nicefrac{1}{2};1) , \mathbb{N} \times (0;1)^3 , 0, \oplus_\text{W}, w \Big)$ with the composition $\oplus_\text{W} : \mathbb{N} \times (0;1)^3 \times  (\nicefrac{1}{2};1)  \longrightarrow  \mathbb{N} \times (0;1)^3$ given by:
\begin{align}
	\begin{aligned}
		(\{n, a , b , c\} , \ F_{m:n} )  \longmapsto  \{ n+1 , a  \frac{1+2F_{m:n}}{3} , b  \frac{1+2F_{m:n}}{3} + a \frac{2(1-F_{m:n})}{3} , c  \frac{1+2F_{m:n}}{3} + b  \frac{2(1-F_{m:n})}{3}\} .
	\end{aligned}
\end{align}

This equation is made evident by using the second equality of Eq. \ref{eq:wmetricfid}. Following the same procedure as before $w(\cdot)$ is given by:
\begin{equation}
	w(\{ n,  a , b , c \}) = 
	\begin{cases}
		a + \frac{2(n-1)}{n^2} b + \frac{2}{n^2} c  &\ ,\text{ if } w(\cdot)\geq \nicefrac{1}{2} \\
		0 \quad &\ ,\text{ otherwise }
	\end{cases}.
\end{equation}

To prove monotonicity is straightfoward. Consider a tree $t$ with corresponding fidelity signature $\{n,a,b,c\} \in \mathbb{N} \times (0,1)^3$. Moreover, consider one additional path with fidelity  given by $F_{m:n} \in  (\nicefrac{1}{2};1)$. We can write the following:
\begin{gather*}
        w(\{n,a,b,c\} \geq w(\{n,a,b,c\} \oplus F_{m:n}) \\
        a + \frac{2(n-1)}{n^2} b + \frac{2}{n^2} c \geq a \cdot \mathcal{F}_{m:n}  + \frac{2n}{(n+1)^2} \left( b \cdot \mathcal{F}_{m:n} + a \cdot \overline{\mathcal{F}}_{m:n} \right)   + \frac{2}{(n+1)^2} \left( c \cdot \mathcal{F}_{m:n} + b \cdot \overline{\mathcal{F}}_{m:n} \right) \\ 
        \Rightarrow a + \frac{2(n-1)}{n^2} b + \frac{2}{n^2} c \geq a \cdot \mathcal{F}_{m:n}  + \frac{2(n-1)}{n^2} \left( b \cdot \mathcal{F}_{m:n} + a \cdot \overline{\mathcal{F}}_{m:n} \right)   + \frac{2}{n^2} \left( c \cdot \mathcal{F}_{m:n} + b \cdot \overline{\mathcal{F}}_{m:n} \right) \\
        a \cdot \overline{\mathcal{F}}_{m:n} \left(1 - \frac{2(n-1)}{n^2} \right) + b \cdot \overline{\mathcal{F}}_{m:n} \left( \frac{2(n-1)}{n^2} - \frac{2}{n^2} \right) + c \cdot \overline{\mathcal{F}}_{m:n} \frac{2}{n^2} \geq 0
\end{gather*}
which is always true for $n\geq 2$, proving therefore monotonicity. As for the label-isotonicity, consider a tree $t$ with a fidelity signature $\{n,a,b,c\} \in \mathbb{N} \times (0,1)^3$ and two different paths with fidelity signatures correspondent to $\sigma_1$, $\sigma_2$ such that $\sigma_1 \preceq \sigma_2$, $i.e.$, $\sigma_1$ has a fidelity given by $F_{m:n}^{(1)}$ and $\sigma_2$ has a fidelity given by $F_{m:n}^{(2)} \leq F_{m:n}^{(2)}$. 
Note that $ F_{m:n}^{(1)} \geq F_{m:n}^{(2)} \Rightarrow \mathcal{F}_{m:n}^{(1)} \geq \mathcal{F}_{m:n}^{(2)} \Rightarrow \overline{\mathcal{F}}_{m:n}^{(1)} \leq \overline{\mathcal{F}}_{m:n}^{(2)}$. Then, we must prove that $w(\{n,a,b,c\} \oplus F_{m:n}^{(1)}) \geq w(\{n,a,b,c\} \oplus F_{m:n}^{(2)})$:

\begin{multline*}
       a \cdot \mathcal{F}_{m:n}^{(1)}  + \frac{2n}{(n+1)^2} \left( b \cdot \mathcal{F}_{m:n}^{(1)} + a \cdot \overline{\mathcal{F}}_{m:n}^{(1)} \right)  + \frac{2}{(n+1)^2} \left( c \cdot \mathcal{F}_{m:n}^{(1)} + b \cdot \overline{\mathcal{F}}_{m:n}^{(1)} \right) \geq  \\
       \geq a \cdot \mathcal{F}_{m:n}^{(2)}  + \frac{2n}{(n+1)^2} \left( b \cdot \mathcal{F}_{m:n}^{(2)} + a \cdot \overline{\mathcal{F}}_{m:n}^{(2)} \right)   + \frac{2}{(n+1)^2} \left( c \cdot \mathcal{F}_{m:n}^{(2)} + b \cdot \overline{\mathcal{F}}_{m:n}^{(2)} \right) 
\end{multline*}
\begin{multline*}
       \Rightarrow a \cdot \mathcal{F}_{m:n}^{(1)} \frac{n^2+1}{(n+1)^2}  +  b \cdot  \mathcal{F}_{m:n}^{(1)}  \frac{2(n-1)}{(n+1)^2} + c \cdot \mathcal{F}_{m:n}^{(1)} \frac{2}{(n+1)^2} \geq \qquad \qquad  \\
       \geq a \cdot \mathcal{F}_{m:n}^{(2)} \frac{n^2+1}{(n+1)^2} + b \cdot \mathcal{F}_{m:n}^{(2)}  \frac{2(n-1)}{(n+1)^2} + c \cdot \mathcal{F}_{m:n}^{(2)} \frac{2}{(n+1)^2}
\end{multline*}
which holds true for every tree, given that $n^2+1$ and $ 2(n-1)$ are always positive quantities, proving therefore the label-isotonicity. This is valid for the distribution of W states with an arbitrary number of qubits.

\color{black}
\section{Further Results on Comparison with existing Multipartite
Routing Algorithms \label{appendix:optimalitysimul}}

In this section we start by presenting additional results for networks that are generated to be similar to internet-like networks:

\begin{figure}[h!]
    \centering
    \subfloat{\includegraphics[width=0.5\linewidth]{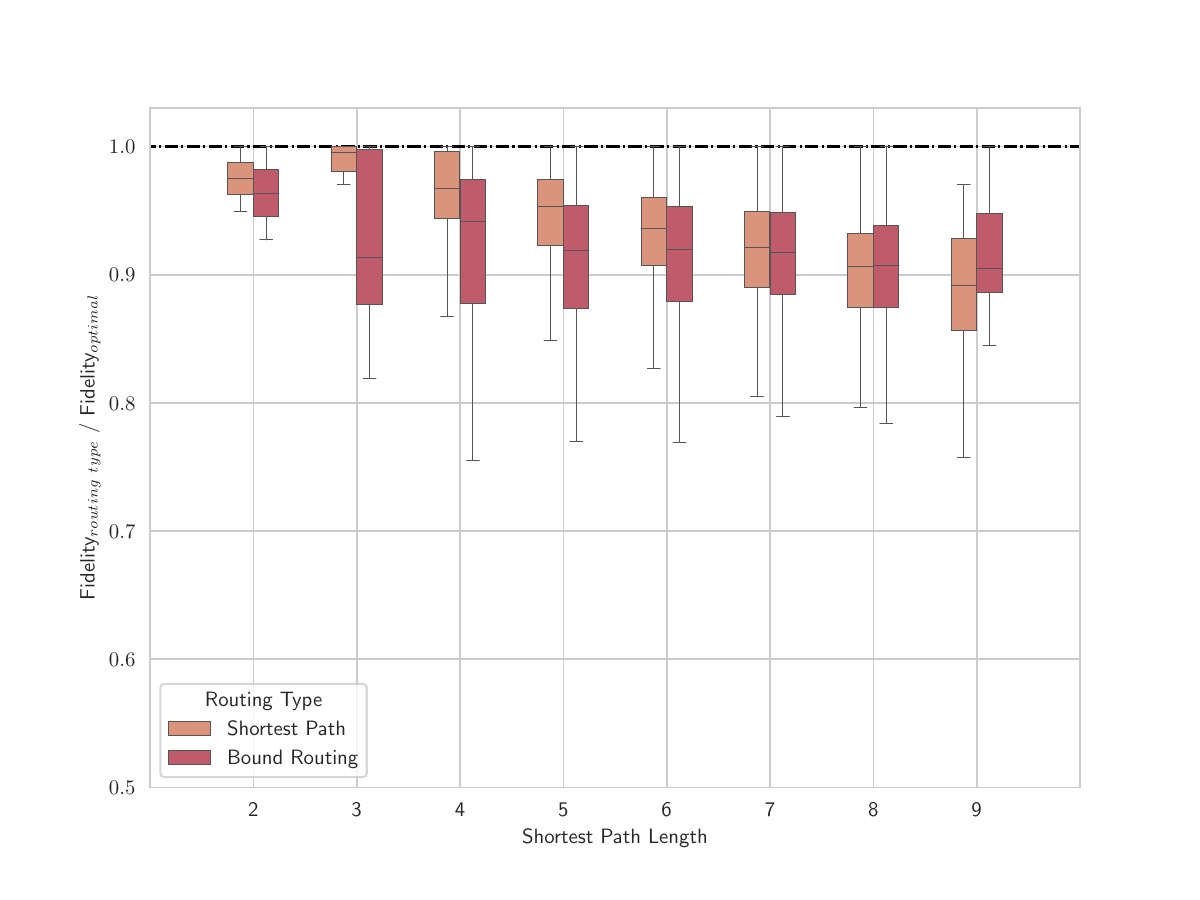}}
    \subfloat{\includegraphics[width=0.5\linewidth]{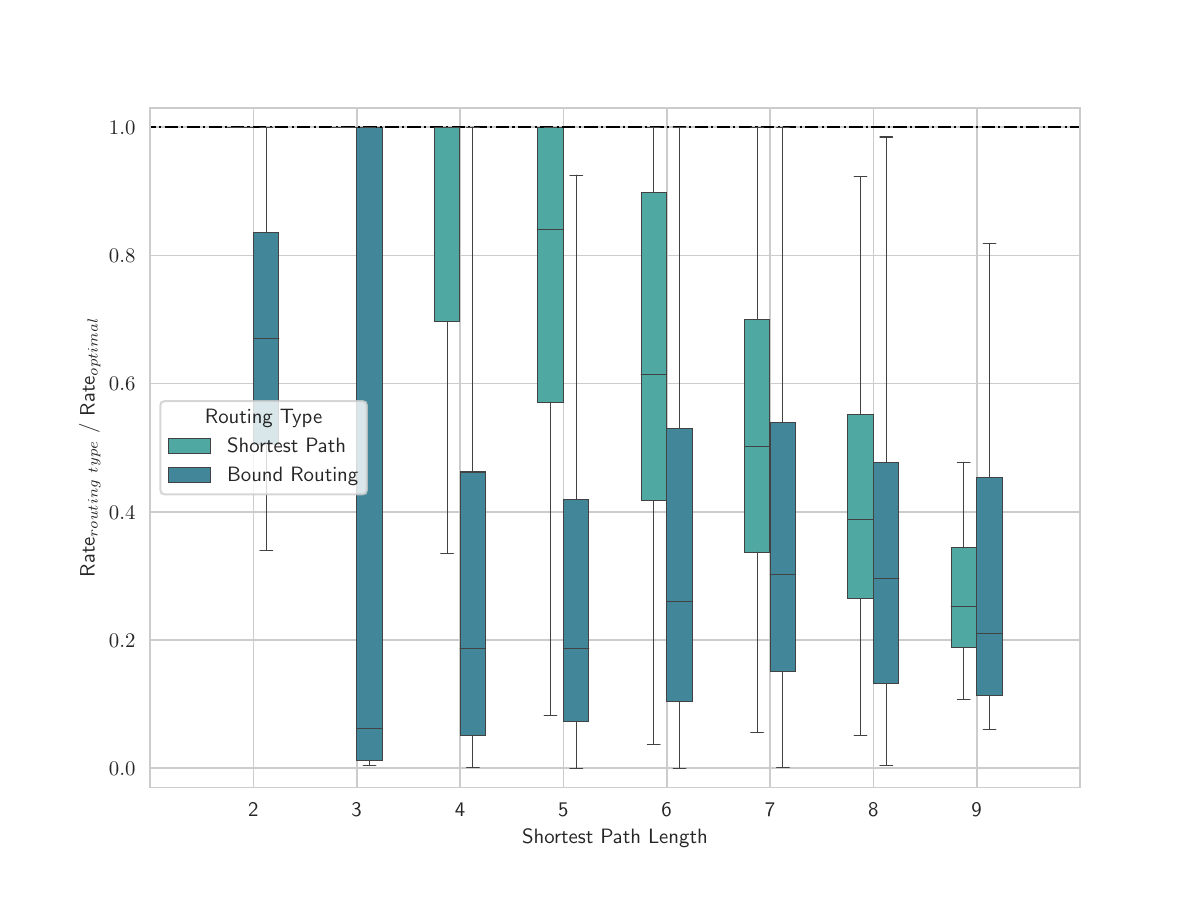}}
    \caption{\color{black}Comparison between optimal routing for a 3-qubit GHZ state, and shortest-path and bound routing for the same state over internet-like networks with 1000 nodes. The parameters utilized are: $\gamma_{min} = 0.9$, $p_\text{min} = 0.5$, $t_\text{min}=1$, $t_\text{max}=100$,  $\sigma_\text{min} = 10^4$, $\sigma_\text{max} =10^5$, $f_\text{trunc}^\text{GHZ}=0.5$, $f_\text{trunc}^\text{path} = 0.72$.\color{black}}
    \label{fig:er}
\end{figure}

As in the case of internet-like networks presented in the main text, we observe that there is no solution found better than our optimal algorithm and, depending on the routing type, one can only hope to achieve fair, but worse, results in terms of both fidelity and rate.

Moreover, it is interesting to consider in what limits do all solutions converge to the same. In terms of routing, when every link's fidelity and memory decoherence time goes to the neutral element ($\gamma \rightarrow 1, \sigma \rightarrow \infty$), every link's communication time is the same and the probability converges to 1, then the routing solution is found by simply finding the scheme where the largest path uses the least amount of links possible:

\begin{equation}
\begin{aligned}
    \gamma_{m:n} &= \prod_{i:j \in m:n} \gamma_{i:j} e^{-t_{i:j}/\sigma_{i:j}} = \prod_{i:j \in m:n} 1 e^{-t_{i:j}/\infty} = 1 \quad \implies \quad f_{GHZ}  = 1 
\end{aligned}
\end{equation}
\begin{equation}
\begin{aligned}
    p_{m:n} &= \prod_{i:j \in m:n} p_{i:j} \rightarrow 1 \quad,\quad t_{m:n} = \sum_{i:j \in m:n} t_{i:j} = n_{m:n}\cdot t   \\
    &\implies \quad t_{GHZ} =  \frac{ 2\cdot \max_{\tau \in \mathcal{T}} \{ t_{c:\tau} \}  }{\prod_{\tau \in \mathcal{T}} p_{c:\tau}} = 2\cdot \max_{\tau \in \mathcal{T}} \{ n_{c:\tau} \}  \cdot t
\end{aligned},
\end{equation}
where $n_{m:n}$ is the number of links of the path.

As means of illustrating this, we performed more simulations, varying the parameter $p_{min}$ when constructing internet-like networks of 1000 nodes, taking the limit where $p_{i:j}$ could only take value 1, while setting $\gamma_{min} = \gamma_{max} = 1$, $\sigma_{min} = \sigma_{max} = \infty$, $t_{min} = t_{max} = 1$. This way, the only difference between paths is their time, which comes in multiples of the same value, depending on their total number of links.

\begin{figure}[h!]
    \centering
    \includegraphics[width=0.7\columnwidth]{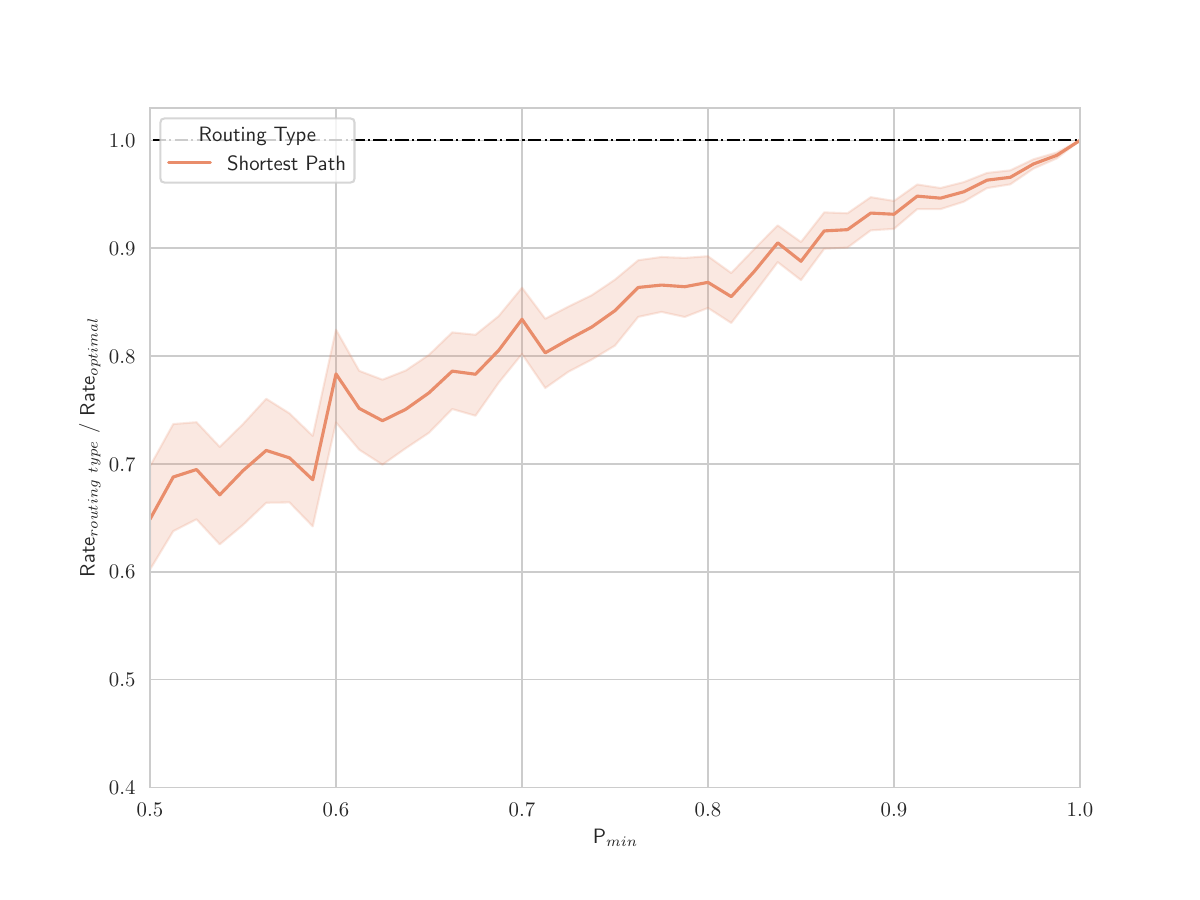}
    \caption{\color{black} Convergence from optimal routing for a 3-qubit GHZ state to shortest-path routing for the same state over internet-like networks with 1000 nodes, by varying the value of $p_{min}$. The parameters utilized are: $\gamma_{min} = 1$, $t_\text{min}=t_\text{max}=1$,  $\sigma_\text{min} = \sigma_\text{max} =\infty$. Each point is the result of 100 simulations over different samples of networks.
    \color{black}}
    \label{fig:convergence}
\end{figure}


\begin{thebibliography}{58}
\providecommand{\natexlab}[1]{#1}
\providecommand{\url}[1]{\texttt{#1}}
\expandafter\ifx\csname urlstyle\endcsname\relax
  \providecommand{\doi}[1]{doi: #1}\else
  \providecommand{\doi}{doi: \begingroup \urlstyle{rm}\Url}\fi

\bibitem[Bennett and Brassard(2014)]{Bennett2014}
Charles~H. Bennett and Gilles Brassard.
\newblock Quantum cryptography: {{Public}} key distribution and coin tossing.
\newblock \emph{Theoretical Computer Science}, 560\penalty0 (P1):\penalty0
  7--11, 2014.
\newblock ISSN 03043975.
\newblock \doi{10.1016/j.tcs.2014.05.025}.

\bibitem[Nurhadi and Syambas(2018)]{Nurhadi2018}
Ali~Ibnun Nurhadi and Nana~Rachmana Syambas.
\newblock Quantum {{Key Distribution}} ({{QKD}}) {{Protocols}}: {{A Survey}}.
\newblock \emph{Proceeding of 2018 4th International Conference on Wireless and
  Telematics, ICWT 2018}, pages 18--22, 2018.
\newblock \doi{10.1109/ICWT.2018.8527822}.

\bibitem[Broadbent et~al.(2009)Broadbent, Fitzsimons, and
  Kashefi]{Broadbent2009}
Anne Broadbent, Joseph Fitzsimons, and Elham Kashefi.
\newblock Universal blind quantum computation.
\newblock \emph{Proceedings - Annual IEEE Symposium on Foundations of Computer
  Science, FOCS}, pages 517--526, 2009.
\newblock ISSN 02725428.
\newblock \doi{10.1109/FOCS.2009.36}.

\bibitem[Chuang(2000)]{Chuang2000}
Isaac Chuang.
\newblock Quantum algorithm for distributed clock synchronization.
\newblock \emph{Physical Review Letters}, 85\penalty0 (9):\penalty0 2006--2009,
  May 2000.
\newblock ISSN 10797114.
\newblock \doi{10.1103/PhysRevLett.85.2006}.

\bibitem[Gottesman et~al.(2011)Gottesman, Jennewein, and Croke]{Gottesman2012}
Daniel Gottesman, Thomas Jennewein, and Sarah Croke.
\newblock Longer-{{Baseline Telescopes Using Quantum Repeaters}}.
\newblock \emph{Physical Review Letters}, 109\penalty0 (7):\penalty0 070503,
  July 2011.
\newblock ISSN 0031-9007.
\newblock \doi{10.1103/PhysRevLett.109.070503}.

\bibitem[Wehner et~al.(2018)Wehner, Elkouss, and Hanson]{Wehner2018}
Stephanie Wehner, David Elkouss, and Ronald Hanson.
\newblock Quantum internet: {{A}} vision for the road ahead.
\newblock \emph{Science}, 362\penalty0 (6412):\penalty0 eaam9288, October 2018.
\newblock ISSN 10959203.
\newblock \doi{10.1126/science.aam9288}.

\bibitem[Pompili et~al.(2021)Pompili, Hermans, Baier, Beukers, Humphreys,
  Schouten, Vermeulen, Tiggelman, {dos Santos Martins}, Dirkse, Wehner, and
  Hanson]{Pompili2021a}
Matteo Pompili, Sophie L.~N. Hermans, Simon Baier, Hans K.~C. Beukers, Peter~C.
  Humphreys, Raymond~N. Schouten, Raymond F.~L. Vermeulen, Marijn~J. Tiggelman,
  L.~{dos Santos Martins}, Bas Dirkse, Stephanie Wehner, and Ronald Hanson.
\newblock Realization of a multinode quantum network of remote solid-state
  qubits.
\newblock \emph{Science}, 372\penalty0 (6539):\penalty0 259--264, April 2021.
\newblock ISSN 0036-8075.
\newblock \doi{10.1126/science.abg1919}.

\bibitem[Alshowkan et~al.(2021)Alshowkan, Williams, Evans, Rao, Simmerman, Lu,
  Lingaraju, Weiner, Marvinney, Pai, Lawrie, Peters, and
  Lukens]{Alshowkan2021a}
Muneer Alshowkan, Brian~P. Williams, Philip~G. Evans, Nageswara~S.V. Rao,
  Emma~M. Simmerman, Hsuan-Hao Lu, Navin~B. Lingaraju, Andrew~M. Weiner,
  Claire~E. Marvinney, Yun-Yi Pai, Benjamin~J. Lawrie, Nicholas~A. Peters, and
  Joseph~M. Lukens.
\newblock Reconfigurable {{Quantum Local Area Network Over Deployed Fiber}}.
\newblock \emph{PRX Quantum}, 2\penalty0 (4):\penalty0 040304, October 2021.
\newblock \doi{10.1103/PRXQuantum.2.040304}.

\bibitem[Munro et~al.(2015)Munro, Azuma, Tamaki, and Nemoto]{Munro2015}
William~J. Munro, Koji Azuma, Kiyoshi Tamaki, and Kae Nemoto.
\newblock Inside {{Quantum Repeaters}}.
\newblock \emph{IEEE Journal of Selected Topics in Quantum Electronics},
  21\penalty0 (3):\penalty0 78--90, May 2015.
\newblock ISSN 1077-260X.
\newblock \doi{10.1109/JSTQE.2015.2392076}.

\bibitem[Caleffi(2017)]{Caleffi2017}
Marcello Caleffi.
\newblock Optimal {{Routing}} for {{Quantum Networks}}.
\newblock \emph{IEEE Access}, 5:\penalty0 22299--22312, 2017.
\newblock ISSN 21693536.
\newblock \doi{10.1109/ACCESS.2017.2763325}.

\bibitem[Chakraborty et~al.(2019)Chakraborty, Rozpedek, Dahlberg, and
  Wehner]{Chakraborty2019a}
Kaushik Chakraborty, Filip Rozpedek, Axel Dahlberg, and Stephanie Wehner.
\newblock Distributed {{Routing}} in a {{Quantum Internet}}, July 2019,
  arXiv:1907.11630.
\newblock \doi{10.48550/arXiv.1907.11630}.

\bibitem[Shi and Qian(2019)]{Shi2019b}
Shouqian Shi and Chen Qian.
\newblock Modeling and {{Designing Routing Protocols}} in {{Quantum Networks}},
  October 2019, arXiv:1909.09329.
\newblock \doi{10.48550/arXiv.1909.09329}.

\bibitem[Li et~al.(2021)Li, Li, Liu, and Cappellaro]{Li2020}
Changhao Li, Tianyi Li, Yi-Xiang~Xiang Liu, and Paola Cappellaro.
\newblock Effective routing design for remote entanglement generation on
  quantum networks.
\newblock \emph{npj Quantum Information}, 7\penalty0 (1):\penalty0 10, December
  2021.
\newblock ISSN 20566387.
\newblock \doi{10.1038/s41534-020-00344-4}.

\bibitem[Dai et~al.(2020)Dai, Peng, and Win]{Dai2020}
Wenhan Dai, Tianyi Peng, and Moe~Z. Win.
\newblock Optimal {{Remote Entanglement Distribution}}.
\newblock \emph{IEEE Journal on Selected Areas in Communications}, 38\penalty0
  (3):\penalty0 540--556, March 2020.
\newblock ISSN 0733-8716.
\newblock \doi{10.1109/JSAC.2020.2969005}.

\bibitem[B{\"a}uml et~al.(2020)B{\"a}uml, Azuma, Kato, and Elkouss]{Bauml2020}
Stefan B{\"a}uml, Koji Azuma, Go~Kato, and David Elkouss.
\newblock Linear programs for entanglement and key distribution in the quantum
  internet.
\newblock \emph{Communications Physics}, 3\penalty0 (1):\penalty0 1--12, 2020.
\newblock ISSN 23993650.
\newblock \doi{10.1038/s42005-020-0318-2}.

\bibitem[Santos et~al.(2023)Santos, Monteiro, Coutinho, and Omar]{Santos2023}
Sara Santos, Francisco~A. Monteiro, Bruno~C. Coutinho, and Yasser Omar.
\newblock Shortest path finding in quantum networks with quasi-linear
  complexity.
\newblock \emph{IEEE Access}, 11:\penalty0 7180--7194, 2023.
\newblock \doi{10.1109/ACCESS.2023.3237997}.

\bibitem[Ren and Hofmann(2012)]{Ren2012}
Changliang Ren and Holger~F. Hofmann.
\newblock Clock synchronization using maximal multipartite entanglement.
\newblock \emph{Physical Review A}, 86\penalty0 (1):\penalty0 014301, July
  2012.
\newblock ISSN 1050-2947.
\newblock \doi{10.1103/PhysRevA.86.014301}.

\bibitem[Khabiboulline et~al.(2019)Khabiboulline, Borregaard, De~Greve, and
  Lukin]{Khabiboulline2019}
E.~T. Khabiboulline, J.~Borregaard, K.~De~Greve, and M.~D. Lukin.
\newblock Quantum-assisted telescope arrays.
\newblock \emph{Physical Review A}, 100\penalty0 (2):\penalty0 022316, August
  2019.
\newblock ISSN 24699934.
\newblock \doi{10.1103/PhysRevA.100.022316}.

\bibitem[Eldredge et~al.(2018)Eldredge, {Foss-Feig}, Gross, Rolston, and
  Gorshkov]{Eldredge2018}
Zachary Eldredge, Michael {Foss-Feig}, Jonathan~A. Gross, Steven~L. Rolston,
  and Alexey~V. Gorshkov.
\newblock Optimal and secure measurement protocols for quantum sensor networks.
\newblock \emph{Physical Review A}, 97\penalty0 (4):\penalty0 042337, April
  2018.
\newblock ISSN 2469-9926.
\newblock \doi{10.1103/PhysRevA.97.042337}.

\bibitem[Qian et~al.(2021)Qian, Bringewatt, Boettcher, Bienias, and
  Gorshkov]{Qian2020}
Timothy Qian, Jacob Bringewatt, Igor Boettcher, Przemyslaw Bienias, and
  Alexey~V. Gorshkov.
\newblock Optimal measurement of field properties with quantum sensor networks.
\newblock \emph{Physical Review A}, 103\penalty0 (3):\penalty0 L030601, March
  2021.
\newblock ISSN 2469-9926.
\newblock \doi{10.1103/PhysRevA.103.L030601}.

\bibitem[Hillery et~al.(1999)Hillery, Bu{\v z}ek, and Berthiaume]{Hillery1999}
Mark Hillery, Vladim{\'i}r Bu{\v z}ek, and Andr{\'e} Berthiaume.
\newblock Quantum secret sharing.
\newblock \emph{Physical Review A - Atomic, Molecular, and Optical Physics},
  59\penalty0 (3):\penalty0 1829--1834, 1999.
\newblock ISSN 10502947.
\newblock \doi{10.1103/PhysRevA.59.1829}.

\bibitem[Zhu et~al.(2015)Zhu, Xu, and Pei]{Zhu2015}
Changhua Zhu, Feihu Xu, and Changxing Pei.
\newblock W-state {{Analyzer}} and {{Multi-party Measurement-device-independent
  Quantum Key Distribution}}.
\newblock \emph{Scientific Reports}, 5\penalty0 (1):\penalty0 17449, December
  2015.
\newblock ISSN 2045-2322.
\newblock \doi{10.1038/srep17449}.

\bibitem[Murta et~al.(2020)Murta, Grasselli, Kampermann, and
  Bru{\ss}]{Murta2020}
Gl{\'a}ucia Murta, Federico Grasselli, Hermann Kampermann, and Dagmar Bru{\ss}.
\newblock Quantum {{Conference Key Agreement}}: {{A Review}}.
\newblock \emph{Advanced Quantum Technologies}, 3\penalty0 (11):\penalty0
  2000025, November 2020.
\newblock ISSN 2511-9044.
\newblock \doi{10.1002/qute.202000025}.

\bibitem[D'Hondt and Panangaden(2006)]{DHondt2004}
Ellie D'Hondt and Prakash Panangaden.
\newblock The {{Computational Power}} of the {{W}} and {{GHZ}} states
\newblock \emph{Quantum Info. Comput.}, 6\penalty0 (2):\penalty0 173–183, Mar 2006.
\newblock ISSN 1533-7146. arXiv:quant-ph/0412177.
\newblock \href{https://doi.org/10.48550/arXiv.quant-ph/0412177}{DOI: 10.48550/arXiv.quant-ph/0412177}. 

\bibitem[Raussendorf and Briegel(2001)]{Raussendorf2001}
Robert Raussendorf and Hans~J Briegel.
\newblock A {{One-Way Quantum Computer}}.
\newblock \emph{Physical Review Letters}, 86\penalty0 (22):\penalty0
  5188--5191, May 2001.
\newblock ISSN 0031-9007.
\newblock \doi{10.1103/PhysRevLett.86.5188}.

\bibitem[Laurenza and Pirandola(2017)]{Laurenza2017}
Riccardo Laurenza and Stefano Pirandola.
\newblock General bounds for sender-receiver capacities in multipoint quantum
  communications.
\newblock \emph{Physical Review A}, 96\penalty0 (3):\penalty0 032318, September
  2017.
\newblock ISSN 2469-9926.
\newblock \doi{10.1103/PhysRevA.96.032318}.

\bibitem[Pirandola(2019{\natexlab{a}})]{Pirandola2019}
Stefano Pirandola.
\newblock End-to-end capacities of a quantum communication network.
\newblock \emph{Communications Physics}, 2\penalty0 (1):\penalty0 51, December
  2019{\natexlab{a}}.
\newblock ISSN 2399-3650.
\newblock \doi{10.1038/s42005-019-0147-3}.

\bibitem[Pirandola(2019{\natexlab{b}})]{Pirandola2019a}
Stefano Pirandola.
\newblock Bounds for multi-end communication over quantum networks.
\newblock \emph{Quantum Science and Technology}, 4\penalty0 (4):\penalty0
  045006, September 2019{\natexlab{b}}.
\newblock ISSN 2058-9565.
\newblock \doi{10.1088/2058-9565/ab3f66}.

\bibitem[Pirandola(2020)]{Pirandola2020}
Stefano Pirandola.
\newblock General upper bound for conferencing keys in arbitrary quantum
  networks.
\newblock \emph{IET Quantum Communication}, 1\penalty0 (1):\penalty0 22--25,
  July 2020.
\newblock ISSN 2632-8925.
\newblock \doi{10.1049/iet-qtc.2020.0006}.

\bibitem[Das et~al.(2021)Das, B{\"a}uml, Winczewski, and Horodecki]{Das2021a}
Siddhartha Das, Stefan B{\"a}uml, Marek Winczewski, and Karol Horodecki.
\newblock Universal {{Limitations}} on {{Quantum Key Distribution}} over a
  {{Network}}.
\newblock \emph{Physical Review X}, 11\penalty0 (4):\penalty0 041016, October
  2021.
\newblock ISSN 2160-3308.
\newblock \doi{10.1103/PhysRevX.11.041016}.

\bibitem[Meignant et~al.(2019)Meignant, Markham, and Grosshans]{Meignant2018}
Cl{\'e}ment Meignant, Damian Markham, and Fr{\'e}d{\'e}ric Grosshans.
\newblock Distributing graph states over arbitrary quantum networks.
\newblock \emph{Physical Review A}, 100\penalty0 (5):\penalty0 052333, November
  2019.
\newblock ISSN 24699934.
\newblock \doi{10.1103/PhysRevA.100.052333}.

\bibitem[Walln{\"o}fer et~al.(2019)Walln{\"o}fer, Pirker, Zwerger, and
  D{\"u}r]{Wallnofer2019}
J.~Walln{\"o}fer, A.~Pirker, M.~Zwerger, and W.~D{\"u}r.
\newblock Multipartite state generation in quantum networks with optimal
  scaling.
\newblock \emph{Scientific Reports}, 9\penalty0 (1):\penalty0 314, December
  2019.
\newblock ISSN 2045-2322.
\newblock \doi{10.1038/s41598-018-36543-5}.

\bibitem[Goodenough et~al.(2021)Goodenough, Elkouss, and
  Wehner]{Goodenough2020}
Kenneth Goodenough, David Elkouss, and Stephanie Wehner.
\newblock Optimizing repeater schemes for the quantum internet.
\newblock \emph{Physical Review A}, 103\penalty0 (3):\penalty0 032610, March
  2021.
\newblock ISSN 2469-9926.
\newblock \doi{10.1103/PhysRevA.103.032610}.

\bibitem[Filippov et~al.(2013)Filippov, Melnikov, and Ziman]{Filippov2013}
Sergey~N. Filippov, Alexey~A. Melnikov, and M{\'a}rio Ziman.
\newblock Dissociation and annihilation of multipartite entanglement structure
  in dissipative quantum dynamics.
\newblock \emph{Physical Review A}, 88\penalty0 (6):\penalty0 062328, December
  2013.
\newblock ISSN 1050-2947.
\newblock \doi{10.1103/PhysRevA.88.062328}.

\bibitem[Sobrinho(2005)]{Sobrinho2005}
J.L. Sobrinho.
\newblock An algebraic theory of dynamic network routing.
\newblock \emph{IEEE/ACM Transactions on Networking}, 13\penalty0 (5):\penalty0
  1160--1173, October 2005.
\newblock ISSN 1063-6692.
\newblock \doi{10.1109/TNET.2005.857111}.

\bibitem[Demeyer et~al.(2013)Demeyer, Goedgebeur, Audenaert, Pickavet, and
  Demeester]{Demeyer2013a}
Sofie Demeyer, Jan Goedgebeur, Pieter Audenaert, Mario Pickavet, and Piet
  Demeester.
\newblock Speeding up {{Martins}}' algorithm for multiple objective shortest
  path problems.
\newblock \emph{4or}, 11\penalty0 (4):\penalty0 323--348, 2013.
\newblock ISSN 16142411.
\newblock \doi{10.1007/s10288-013-0232-5}.

\bibitem[Brand et~al.(2020)Brand, Coopmans, and Elkouss]{Brand2020}
Sebastiaan Brand, Tim Coopmans, and David Elkouss.
\newblock Efficient {{Computation}} of the {{Waiting Time}} and {{Fidelity}} in
  {{Quantum Repeater Chains}}.
\newblock \emph{IEEE Journal on Selected Areas in Communications}, 38\penalty0
  (3):\penalty0 619--639, March 2020.
\newblock ISSN 0733-8716.
\newblock \doi{10.1109/JSAC.2020.2969037}.

\bibitem[Werner(1989)]{Werner1989}
Reinhard~F. Werner.
\newblock Quantum states with {{Einstein-Podolsky-Rosen}} correlations
  admitting a hidden-variable model.
\newblock \emph{Physical Review A}, 40\penalty0 (8):\penalty0 4277--4281, 1989.
\newblock ISSN 10502947.
\newblock \doi{10.1103/PhysRevA.40.4277}.

\bibitem[Hein et~al.(2006)Hein, D{\"u}r, Eisert, Raussendorf, den Nest, and
  Briegel]{Hein2006}
M.~Hein, W.~D{\"u}r, J.~Eisert, R.~Raussendorf, M.~Van den Nest, and H.~J.
  Briegel.
\newblock Entanglement in {{Graph States}} and its {{Applications}}.
\newblock \emph{Proceedings of the International School of Physics "Enrico
  Fermi"}, 162:\penalty0 115--218, February 2006.
\newblock ISSN 0074784X.
\newblock \doi{10.3254/978-1-61499-018-5-115}.

\bibitem[D{\"u}r and Briegel(2007)]{Dur2007}
W.~D{\"u}r and H.~J. Briegel.
\newblock Entanglement purification and quantum error correction.
\newblock \emph{Reports on Progress in Physics}, 70\penalty0 (8):\penalty0
  1381--1424, 2007.
\newblock ISSN 00344885.
\newblock \doi{10.1088/0034-4885/70/8/R03}.

\bibitem[neng Guo et~al.(2017)neng Guo, long Tian, Zeng, and da~Li]{Guo2017}
You neng Guo, Qing long Tian, Ke~Zeng, and Zheng da~Li.
\newblock Quantum coherence of two-qubit over quantum channels with memory.
\newblock \emph{Quantum Information Processing}, 16\penalty0 (12):\penalty0
  1--18, 2017.
\newblock ISSN 15700755.
\newblock \doi{10.1007/s11128-017-1749-x}.

\bibitem[Kamin et~al.(2022)Kamin, Shchukin, Schmidt, and {van
  Loock}]{Kamin2022a}
Lars Kamin, Evgeny Shchukin, Frank Schmidt, and Peter {van Loock}.
\newblock Exact rate analysis for quantum repeaters with imperfect memories and
  entanglement swapping as soon as possible, March 2022, arXiv:2203.10318.
\newblock \doi{10.48550/arXiv.2203.10318}.

\bibitem[Martins(1984)]{Martins1984}
Ernesto Queir{\'o}s~Vieira Martins.
\newblock On a multicriteria shortest path problem.
\newblock \emph{European Journal of Operational Research}, 16\penalty0
  (2):\penalty0 236--245, 1984.
\newblock ISSN 03772217.
\newblock \doi{10.1016/0377-2217(84)90077-8}.

\bibitem[Sobrinho(2003)]{Sobrinho2003}
Jo{\~a}o~Lu{\'i}s Sobrinho.
\newblock Network {{Routing}} with {{Path Vector Protocols}}: {{Theory}} and
  {{Applications}}.
\newblock \emph{Computer Communication Review}, 33\penalty0 (4):\penalty0
  49--60, 2003.
\newblock ISSN 01464833.
\newblock \doi{10.1145/863955.863963}.

\bibitem[Barab{\'a}si and P{\'o}sfai(2016)]{Barabasi2016}
Albert-L{\'a}szl{\'o} Barab{\'a}si and M{\'a}rton P{\'o}sfai.
\newblock \emph{Network Science}.
\newblock {Cambridge University Press}, {Cambridge}, 2016.
\newblock ISBN 978-1-107-07626-6 1-107-07626-9.

\bibitem[Dorogovtsev et~al.(2008)Dorogovtsev, Goltsev, and
  Mendes]{Dorogovtsev2008}
S.~N. Dorogovtsev, A.~V. Goltsev, and J.~F.F. Mendes.
\newblock Critical phenomena in complex networks.
\newblock \emph{Reviews of Modern Physics}, 80\penalty0 (4):\penalty0
  1275--1335, 2008.
\newblock ISSN 00346861.
\newblock \doi{10.1103/RevModPhys.80.1275}.

\bibitem[Ellis et~al.(2007)Ellis, Martin, and Yan]{Ellis2007}
Robert~B. Ellis, Jeremy~L. Martin, and Catherine Yan.
\newblock Random geometric graph diameter in the unit ball.
\newblock \emph{Algorithmica (New York)}, 47\penalty0 (4):\penalty0 421--438,
  2007.
\newblock ISSN 01784617.
\newblock \doi{10.1007/s00453-006-0172-y}.

\bibitem[Dall and Christensen(2002)]{Dall2002}
Jesper Dall and Michael Christensen.
\newblock Random geometric graphs.
\newblock \emph{Physical Review E - Statistical Physics, Plasmas, Fluids, and
  Related Interdisciplinary Topics}, 66\penalty0 (1), 2002.
\newblock ISSN 1063651X.
\newblock \doi{10.1103/PhysRevE.66.016121}.

\bibitem[Inagaki et~al.(2013)Inagaki, Matsuda, Tadanaga, Asobe, and
  Takesue]{Inagaki2013}
Takahiro Inagaki, Nobuyuki Matsuda, Osamu Tadanaga, Masaki Asobe, and Hiroki
  Takesue.
\newblock Entanglement distribution over 300 km of fiber.
\newblock \emph{Optics Express}, 21\penalty0 (20):\penalty0 23241, 2013.
\newblock ISSN 1094-4087.
\newblock \doi{10.1364/oe.21.023241}.

\bibitem[Coutinho et~al.(2022)Coutinho, Munro, Nemoto, and Omar]{Coutinho2022}
Bruno~Coelho Coutinho, William~John Munro, Kae Nemoto, and Yasser Omar.
\newblock Robustness of noisy quantum networks.
\newblock \emph{Communications Physics}, 5\penalty0 (1):\penalty0 1--9, April
  2022.
\newblock ISSN 2399-3650.
\newblock \doi{10.1038/s42005-022-00866-7}.

\bibitem[Avis et~al.(2022)Avis, Rozp{\k{e}}dek, and Wehner]{Avis2022a}
Guus Avis, Filip Rozp{\k{e}}dek, and Stephanie Wehner.
\newblock Analysis of {{Multipartite Entanglement Distribution}} using a
  {{Central Quantum-Network Node}}, March 2022, arXiv:2203.05517.
\newblock \doi{10.48550/arXiv.2203.05517}.

\bibitem[Walln{\"o}fer et~al.(2016)Walln{\"o}fer, Zwerger, Muschik, Sangouard,
  and D{\"u}r]{Wallnofer2016}
J.~Walln{\"o}fer, M.~Zwerger, C.~Muschik, N.~Sangouard, and W.~D{\"u}r.
\newblock Two-dimensional quantum repeaters.
\newblock \emph{Physical Review A}, 94\penalty0 (5):\penalty0 1--12, 2016.
\newblock ISSN 24699934.
\newblock \doi{10.1103/PhysRevA.94.052307}.

\bibitem[Satoh et~al.(2016)Satoh, Ishizaki, Nagayama, and Van~Meter]{Satoh2016}
Takahiko Satoh, Kaori Ishizaki, Shota Nagayama, and Rodney Van~Meter.
\newblock Analysis of quantum network coding for realistic repeater networks.
\newblock \emph{Physical Review A}, 93\penalty0 (3):\penalty0 1--10, 2016.
\newblock ISSN 24699934.
\newblock \doi{10.1103/PhysRevA.93.032302}.

\bibitem[Sekatski et~al.(2019)Sekatski, W{\"o}lk, and D{\"u}r]{Sekatski2019}
Pavel Sekatski, Sabine W{\"o}lk, and Wolfgang D{\"u}r.
\newblock Optimal distributed sensing in noisy environments.
\newblock \emph{Physical Review Research}, 2\penalty0 (2):\penalty0 1--8, May
  2019.
\newblock \doi{10.1103/PhysRevResearch.2.023052}.

\bibitem[Shettell et~al.(2021)Shettell, Munro, Markham, and
  Nemoto]{Shettell2021}
Nathan Shettell, William~J. Munro, Damian Markham, and Kae Nemoto.
\newblock Practical limits of error correction for quantum metrology.
\newblock \emph{New Journal of Physics}, 23\penalty0 (4):\penalty0 043038,
  April 2021.
\newblock ISSN 1367-2630.
\newblock \doi{10.1088/1367-2630/abf533}.

\bibitem[Wang(2008)]{Wang2008}
X.~Wang.
\newblock \emph{Exact Algorithms for {{Steiner}} Tree Problem}.
\newblock 2008.
\newblock ISBN 978-90-365-2660-9.
\newblock \doi{10.3990/1.9789036526609}.

\bibitem[Robins and Zelikovsky(2005)]{Robins2005}
Gabriel Robins and Alexander Zelikovsky.
\newblock Tighter {{Bounds}} for {{Graph Steiner Tree Approximation}}.
\newblock \emph{SIAM Journal on Discrete Mathematics}, 19\penalty0
  (1):\penalty0 122--134, January 2005.
\newblock ISSN 0895-4801.
\newblock \doi{10.1137/S0895480101393155}.

\bibitem[D{\"u}r et~al.(2000)D{\"u}r, Vidal, and Cirac]{Dur2000}
W.~D{\"u}r, G~Vidal, and J~I Cirac.
\newblock Three qubits can be entangled in two inequivalent ways.
\newblock \emph{Physical Review A}, 62\penalty0 (6):\penalty0 062314, November
  2000.
\newblock ISSN 1050-2947.
\newblock \doi{10.1103/PhysRevA.62.062314}.

\end{thebibliography}
\end{document}